\documentclass[12pt,dvipsnames]{article}

\usepackage[top=80pt,bottom=85pt,left=85pt,right=85pt]{geometry}

\usepackage{authblk}

\usepackage[utf8]{inputenc}
\usepackage[T1]{fontenc}
\usepackage[english]{babel}
\usepackage{microtype}

\usepackage{amsmath}
\usepackage{amssymb}
\usepackage{physics}
\usepackage{dsfont}
\usepackage{amsthm}
\usepackage{mathrsfs}
\numberwithin{equation}{section}

\usepackage[dvipsnames]{xcolor}
\usepackage{tikz}
\usetikzlibrary{arrows, arrows.meta, positioning, calc, decorations.markings}
\usepackage{tikz-cd} % Added for the commutative diagram
\usepackage{subcaption}
\usepackage{float}

\usepackage{booktabs}
\usepackage{tabularx}
\usepackage{enumitem}

\usepackage[pdftex,colorlinks=true]{hyperref}
\hypersetup{urlcolor=RedOrange, citecolor=BurntOrange, linkcolor=OliveGreen}
\usepackage{cleveref}

\DeclareCaptionFormat{custom}{#1#2#3}

\newcommand{\bs}{\boldsymbol}
\renewcommand{\tr}{\mathrm{Tr}}

\newcommand{\rep}[1]{\rho_{#1}} 
\newcommand{\mult}[1]{\mathcal{V}_{#1}} 
\newcommand{\tensor}{\otimes}

\begin{document}
% --- Title and Authors ---
\title{\Large \textbf{Non-toric 5d SCFTs from Reid's Pagoda}}

% \author{
%     Andr\'{e}s Collinucci$^{1}$, 
%     Fabrizio Del Monte$^{2}$, \\
%     Mario De Marco$^{1}$, 
%     Marina Moleti$^{3,4}$,
%     and Roberto Valandro$^{3,4}$
% }

% \date{%
%     \small $^{1}$Service de Physique de l'Univers, Champs et Gravitation, Universit\'{e} Libre de Bruxelles\\
%     $^{2}$School of Mathematics, University of Birmingham\\
%     $^{3}$Dipartimento di Fisica, Universit\`{a} di Trieste\\
%     $^{4}$INFN, Sezione di Trieste\\[2ex]
% }

\author[1]{Andr\'{e}s Collinucci}
\author[2]{Fabrizio Del Monte}
\author[1]{Mario De Marco}
\author[3]{Marina Moleti}
\author[4]{Roberto Valandro}

\affil[1]{Physique Th\'eorique et Math\'ematique and International Solvay Institutes\protect\\
Universit\'e Libre de Bruxelles, C.P. 231, 1050 Brussels, Belgium}
\affil[2]{School of Mathematics, University of Birmingham, Watson Building, Edgbaston, Birmingham B15
2TT, UK}
\affil[3]{SISSA and INFN, Via Bonomea 265, I-34136 Trieste, Italy}
\affil[4]{Dipartimento di Fisica, Università di Trieste, Strada Costiera 11, I-34151 Trieste, Italy\protect\\
and INFN, Sezione di Trieste, Via Valerio 2, I-34127 Trieste, Italy}
\date{}
\maketitle
\thispagestyle{empty}
\begin{center}
\texttt{collinucci.phys@gmail.com, f.delmonte.mp@gmail.com,  mario.de.marco@ulb.be, mmoleti@sissa.it}\\
\texttt{roberto.valandro@ts.infn.it}
\end{center}
\vskip 1.0 cm
\begin{abstract}
We construct new families of non-toric 5d SCFTs via abelian orbifolds of the Reid Pagoda, including a surprising infinite family of rank-1 theories, that evade all known classifications. Using the McKay correspondence, we derive their BPS quivers and superpotentials. The hallmark of these theories is a novel sector we dub \textit{Pagoda matter}, whose vacuum expectation values obstruct the K\"ahler moduli. This mechanism freezes the gauge coupling to infinite value, precluding a weak-coupling limit and rendering the theories intrinsically strongly coupled. Finally, we interpret these results as 5d SCFTs deformed by non-constant flavor backgrounds.
\end{abstract}

\newpage
\tableofcontents
% ==========================================
% SECTION 1: INTRODUCTION
% ==========================================
\section{Introduction}
\setcounter{page}{1}
% OLD: Existing classificatons of geometrically engineered 5d Superconformal Field Theories (SCFTs) 
% OLD: rely either on the availability of toric models or on the existence of a weakly-coupled 
% OLD: gauge theory regime somewhere in the moduli space. In this paper, we show that 
% OLD: there is a large undiscovered sector...

The classification of 5d superconformal field theories (SCFTs) has historically rested on two pillars: the availability of toric models (and their associated $(p,q)$-web descriptions \cite{Franco:2005rj, Hanany:2005ve}) and the existence of a weakly-coupled gauge theory regime somewhere on the moduli space. In this work, we demonstrate that both pillars are optional. We uncover a vast landscape of 5d SCFTs that are intrinsically non-toric and, more strikingly, possess an exotic type of matter which, when activated, drives the theory away \emph{from the weak-coupling limit}.
These theories reside at canonical singularities that defy standard classification schemes, such as those found in \cite{Seiberg:1996bd, Intriligator:1997pq, Jefferson:2018irk} among many other works, and are characterized by a novel matter sector we refer to as \textit{Pagoda matter}.

The core of our construction is a non-toric generalization of the conifold known as the Reid Pagoda \cite{Reid1983_MinMod3folds}, defined by the hypersurface $uv = z^2 - w^{2k}$ in $\mathbb{C}^4$.  While the parent Pagoda describes a rank-zero theory with a single floppable $\mathbb{P}^1$, we systematically generate rank-$N$ theories by taking abelian orbifolds of the form $(Y_{\text{Pagoda}})/H$, with $Y_{\text{Pagoda}}$ the Reid Pagoda singularity. Using the McKay correspondence, we provide the complete BPS quiver and superpotential for these theories.

The physical "punch-line" of these geometries is a mechanism we call the \emph{freezing of the gauge coupling}. In standard 5d SCFTs, the inverse bare coupling $m_0$ corresponds to a tunable K\"{a}hler modulus—the volume of a base curve. In our Pagoda-orbifold constructions, however, the vacuum expectation values of the "Pagoda matter" sector act as an obstruction. Specifically, activating these VEVs forces the K\"{a}hler modulus to vanish ($m_0 \to 0$), effectively removing the weak-coupling "buffer." As a result, the theory is intrinsically strongly coupled at the origin of the Coulomb branch and cannot be continuously deformed to a perturbative non-Abelian gauge theory.

Finally, we provide a robust physical interpretation of this phenomenon by viewing these 5d SCFTs as deformations of toric orbifolds (such as local $\mathbb{F}_2$) by a \emph{non-constant flavor background}. By promoting the global flavor symmetry parameters to position-dependent Higgs fields $\Phi(w)$, we show that the geometry now has only an isolated point-like singularity supporting a localized sector of fluctuations. We explain how the exotic Pagoda matter is "trapped" by the geometry and elucidating the observed obstruction of the K\"{a}hler moduli. We bolster our claim by supplying a 3d mirror symmetry analysis of D2-brane probes of our geometries, based on work in progress \cite{Collinucci:D2branes}. 

These results open up several new directions for future investigation. Among the most interesting are the analysis of wall-crossing and the stability of BPS states, aimed at determining the BPS spectrum of the theories presented here, and the extension of the procedure developed in this paper to higher-length flops \cite{Collinucci:2022rii}, i.e., isolated threefold singularities generalizing Reid’s Pagoda. In the latter case, preliminary results suggest that infinite families of higher-rank 5d SCFTs emerge from abelian orbifolds of the Laufer length-two flop.

\textbf{Outline.} Section \ref{sec:PagodaCon} reviews the Reid Pagoda and its quiver description. Section \ref{sec:OrbifConstrPag} develops the orbifold construction at the level of quivers and superpotentials. Section \ref{sec:OrbGeom} analyses the resulting moduli spaces and their compact 4-cycles. Section \ref{sec:HiggsPagoda} derives the central physical result: the freezing of the gauge coupling at infinity. Section \ref{sec:Physics} offers a physical interpretation of the construction in terms of a non-constant flavour background, first from the viewpoint of D2-branes and then directly in 5d. Finally, Section \ref{sec:GeneralOrb} extends the construction to more general abelian orbifolds, giving explicit BPS quivers with superpotentials for new 5d theories with rank anywhere from zero to an arbitrary high number. In Appendix \ref{app:BPSQuivers} we briefly recall all the necessary information about BPS quivers of 5d geometrically engineered QFTs.

\section{The Parent Theory: Pagoda Flops}\label{sec:PagodaCon}
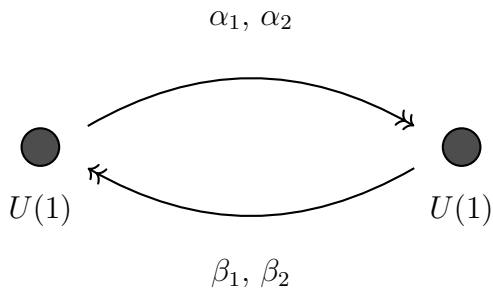
\begin{figure}[htbp]
\centering
\begin{tikzpicture}[
  x=1.4cm,y=1.4cm,
  place/.style={circle,draw=black,fill=black!70,thick,inner sep=0pt,minimum size=5mm}
]
\node at (-2,0) [place] (L) {};
\node at ( 2,0) [place] (R) {};
\node at (-2,-.6) {$U(1)$};
\node at ( 2,-.6) {$U(1)$};
\draw[thick, ->>] (-1.55,0.20) to[bend left] (1.55,0.20);
\node at (0,1.2) {$\alpha_{1},\,\alpha_{2}$};
\draw[thick, ->>] (1.55,-0.20) to[bend left] (-1.55,-0.20);
\node at (0,-1.2) {$\beta_{1},\,\beta_{2}$};
\end{tikzpicture}
\caption{Abelian Klebanov-Witten quiver.}
\label{fig:KW}
\end{figure}

The first ingredient we need is a construction of threefolds that behave very similarly to the conifold, usually described as the equation in $\mathbb{C}^4$
\begin{equation}
X_{\rm conifold}: \quad u v = z^2+w^2\,,
\end{equation}
which admits a small resolution $\tilde X_{\rm conifold}$, such that the smooth threefold is the total space of the normal bundle to the flopping curve, i.e.:
\begin{equation}
\tilde X_{\rm conifold} \cong \mathcal{O}(-1) \oplus \mathcal{O}(-1)_{\mathbb{P}^1}\,.
\end{equation}
This resolved space admits a description as the moduli space of the Abelian Klebanov-Witten quiver \cite{Klebanov:1998hh} (see Figure \ref{fig:KW}).

For this Abelian case, the superpotential is trivial, and the moduli space is toric. The quiver (with superpotential) can be physically interpreted either as the $d=4, \mathcal{N}=1$ theory on a probe D3-brane, or as the supersymmetric quantum mechanics of a probe D0-brane. In this paper, we will always choose the D0-brane interpretation, which will be later generalized beyond the abelian case through the notion of \emph{BPS quivers} \cite{Cecotti2011,Alim2013,Closset2022}. Therefore, all arrows appearing in our quivers are to be interpreted as chiral multiplets for \emph{$d=1, \mathcal{N}=4$ quiver quantum mechanics}.
M-theory on the conifold is believed to give rise to a 5d (trivial) SCFT with a free hypermultiplet, realized by an M2-brane wrapping the exceptional $\mathbb{P}^1$. The only non-vanishing Gopakumar-Vafa invariant is $n_{d=1}^{g=0} = 1$, which matches this expectation.

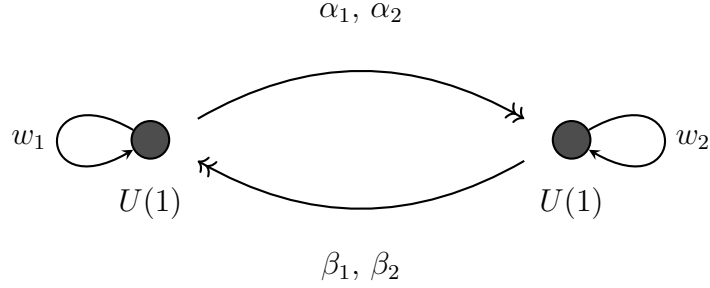
\begin{figure}[htbp]
\centering
\begin{tikzpicture}[
  x=1.4cm,y=1.4cm,
  place/.style={circle,draw=black,fill=black!70,thick,inner sep=0pt,minimum size=5mm}
]
\node at (-2,0) [place] (L) {};
\node at ( 2,0) [place] (R) {};
\node at (-2,-.6) {$U(1)$};
\node at ( 2,-.6) {$U(1)$};
\draw[thick,-stealth,looseness=15,out=150,in=210] (L) to node[left]  {$w_{1}$} (L);
\draw[thick,-stealth,looseness=15,out=30,in=-30]  (R) to node[right] {$w_{2}$} (R);
\draw[thick,->>]  (-1.55,0.20) to[bend left]  (1.55,0.20);
\node at (0,1.2) {$\alpha_{1},\,\alpha_{2}$};
\draw[thick,->>] (1.55,-0.20) to[bend left] (-1.55,-0.20);
\node at (0,-1.2) {$\beta_{1},\,\beta_{2}$};
\end{tikzpicture}
\caption{The D0-brane quiver of the Reid Pagoda.}
\label{fig:Reidspagoda}
\end{figure}

Our main object of interest will be the class of local Calabi-Yau threefolds known as Reid's \emph{Pagodas} \cite{Reid1983_MinMod3folds}, given by
\begin{equation} \label{kpagoda}
X_{\rm k-Pagoda}: \quad uv = z^2-w^{2 k}\,.
\end{equation}
They are an infinite family indexed by an integer $k$, with the $k=1$ case being the conifold. They too, admit a small resolution that gives rise to a single floppable $\mathbb{P}^1$. Much like the conifold, we can perform a small resolution of this via the following standard procedure: We enhance the ambient space from $\mathbb{C}^4$ to a product space $\mathbb{C}^4 \times \mathbb{P}^1$, and we impose the following system of constraints:
\begin{equation} \label{eq:pagodasmallres}
\mathcal{R}\cdot \begin{pmatrix}
    s_1\\s_2
\end{pmatrix} = 0 \qquad \subset \mathbb{C}^4_{u, v, z, w} \times \mathbb{P}^1_{[s_1: s_2]}\,,
\end{equation}
whereby
\begin{equation}
    \mathcal{R}:= \begin{pmatrix}
        u&z+w^k\\z-w^k&v
    \end{pmatrix}
\end{equation}
is a two-by-two matrix whose determinant is the defining equation for the Pagoda hypersurface. Hence, along the hypersurface rk$(\mathcal{R})\leq 1$, and the matrix-valued constraint \eqref{eq:pagodasmallres} fixes a point in $\mathbb{P}^1$. Whenever $\mathcal{R}=0$ strictly, then \eqref{eq:pagodasmallres} is trivially satisfied, and a whole $\mathbb{P}^1$ is `freed up'. Similarly to the conifold, this resolution can be described as the moduli space of the quiver in Figure \ref{fig:Reidspagoda}, with $U(1) \times U(1)$ gauge group.

The theory exhibits a manifest $SU(2)$ flavor symmetry acting on the bifundamental fields. The pairs $(\alpha_1, \alpha_2)$ and $(\beta_1, \beta_2)$ each transform as doublets under this symmetry. To write the superpotential compactly and manifestly invariant, we introduce the $SU(2)$-invariant symplectic contraction: 
\begin{equation}
    \boldsymbol{\alpha} \cdot \boldsymbol{\beta} := \epsilon^{ij} \alpha_i \beta_j 
    = \alpha_1 \beta_2 - \alpha_2 \beta_1 \,.
\end{equation}
With this notation, the superpotential takes the form:
\begin{equation} \label{eq:PagodaW}
  \mathcal{W} = \text{Tr} \left[ w_1 \left(\boldsymbol{\alpha} \cdot \boldsymbol{\beta} + \frac{1}{k+1}w_1^{k}\right) + w_2 \left(\boldsymbol{\beta} \cdot \boldsymbol{\alpha} - \frac{1}{k+1}w_2^{k}\right) \right] \,.
\end{equation}

The sum of the two terms linear in the $w_i$'s give the usual superpotential that would have led to $\mathbb{C}^2/\mathbb{Z}_2 \times \mathbb{C}$. For $k=1$, the higher order terms correspond to the mass terms used by Klebanov and Witten.

One easily recovers the singular geometry by defining the bilinear invariants. The F-term equations with respect to $w_1$ and $w_2$ impose:
\begin{equation}
    \boldsymbol{\alpha} \cdot \boldsymbol{\beta} + w_1^k = 0 \quad \text{and} \quad -(\boldsymbol{\beta} \cdot \boldsymbol{\alpha}) + w_2^k = 0 \,.
\end{equation}
Using the property that $\boldsymbol{\beta} \cdot \boldsymbol{\alpha} = - \boldsymbol{\alpha} \cdot \boldsymbol{\beta}$ in the abelian case, this implies:
\begin{equation}
    w_1^k = -(\boldsymbol{\alpha} \cdot \boldsymbol{\beta}) \quad \text{and} \quad w_2^k = \boldsymbol{\beta} \cdot \boldsymbol{\alpha} = -(\boldsymbol{\alpha} \cdot \boldsymbol{\beta})\,.
\end{equation}
The F-terms of $\boldsymbol{\alpha}$ and $\boldsymbol{\beta}$ imply that $w_1 = w_2 \equiv w$, so we simply write:
\begin{equation} \label{eq:Fterm}
   \alpha_1 \beta_2 - \alpha_2 \beta_1 = -w^k \,.
\end{equation}
We define the remaining affine coordinates as the symmetric combinations of the fields (with a factor of 2 for convenience):
\begin{equation}
\label{eq:pagodacordstoquiver}
    u \equiv 2\alpha_1 \beta_1 \,, \quad v \equiv 2\alpha_2 \beta_2 \,, \quad z \equiv \alpha_1 \beta_2 + \alpha_2 \beta_1 \,.
\end{equation}
To find the relation between them, we square the $z$ coordinate and use the F-term constraint:
\begin{equation}
    z^2 - w^{2k} = (\alpha_1 \beta_2 + \alpha_2 \beta_1)^2 - (\alpha_1 \beta_2 - \alpha_2 \beta_1)^2 = 4 (\alpha_1 \beta_2)(\alpha_2 \beta_1) \,.
\end{equation}
Comparing this to the product $uv$, we see:
\begin{equation}
    uv = (2\alpha_1 \beta_1)(2\alpha_2 \beta_2) = 4 \alpha_1 \beta_2 \alpha_2 \beta_1 \,.
\end{equation}
Thus, we recover the Reid Pagoda equation exactly:
\begin{equation}
    uv - z^2 + w^{2k} = 0 \,.
\end{equation}

The key point of interest about the Pagoda is that, despite having a single exceptional (floppable) $\mathbb{P}^1$, its GV invariant is $n_{d=1}^{g=0} = k$. This suggests a flavor symmetry of $USp(2)$ rather than $USp(2k)$.

The fact that this geometry has only a single $\mathbb{P}^1$ an the striking consequence, that only one real mass parameter is available. The corresponding K\"ahler modulus gives an equal mass to all $k$ matter fields, without possibility of tuning them independently. The QFT explanation for such a phenomenon is not known to us. One possible mechanism for this might be a discrete $S_k$ gauge symmetry acting on the matter fields.

This peculiar matter sector is central to our construction of non-toric 5d SCFTs.

% SECTION 3: THE ORBIFOLD CONSTRUCTION
% ==========================================
\section{The Orbifold Construction}\label{sec:OrbifConstrPag}
In the previous subsections, we constructed rank-zero theories. By orbifolding these geometries, however, four-cycles will inevitably be generated, taking us to rank-one theories. As before, the geometry will emerge from a quiver, that will be obtained through a powerful mathematical construction known as the McKay correspondence. These (not necessarily abelian) quivers are known as BPS quivers, and describe the supersymmetric quantum mechanics on the worldline of BPS states formed by D0-D2-D4 bound states on the orbifold, see Appendix \ref{app:BPSQuivers} for more details.

% The bridge between the geometry and the quiver is the Derived McKay Correspondence \cite{Bridgeland2001}. For an orbifold $Y/H$, there is an equivalence of derived categories $D_c(X)\cong D_c^H(Y)$, where $X$ is the crepant resolution and $D_c^H(Y)$ is the derived category of $H$-equivariant coherent sheaves. Practically, this implies that the fractional branes of the theory correspond to the irreducible representations of the orbifold group $H$, and the quiver can be derived via group theory.

\subsection{Orbifold quivers and the McKay correspondence}\label{sec:McKay}
We construct the orbifold theory following the logic of the McKay correspondence. One can think of what we are about to recall as a vast generalization of the original work of Douglas and Moore \cite{Douglas:1996sw} describing D-branes probing orbifolds $\mathbb{C}^2/\mathbb{Z}_N$. %\mdm{se mettessimo $\mathbb C^2/Z_N \times \mathbb C$? Douglas e Moore danno la prescrizione per il superpotenziale (che a mio perere e` quello del threefold con gli aggiunti, non mi pare si possa scrivere un superpotenziale per il quiver solo di $\mathbb C^2/Z_N$)}
In its general formulation \cite{Bridgeland2001}, the McKay correspondence is formulated as an equivalence between BPS states on a CY3 $Y$ and on the resolution of its orbifold $Y/H$ by a finite group $H$\footnote{For the mathematically oriented reader, this is an equivalence between the bounded derived category of compactly supported coherent sheaves on the resolved orbifold, and the equivariant derived category on the original space, $D_c\left(\widehat{Y/H}\right)\cong D^H_c(Y)$, see \cite{Aspinwall2009} for a review and \cite{DelMonte2023,DelMonte2024,Dell2025,Bridgeland2024} for recent developments in the context of 5d BPS spectra.}. In practice, this gives an explicit construction of fractional branes of the theory in terms of irreducible representations of the orbifold group $H$. Likewise, the quiver and superpotential will be also derived via group theory.

% While the classical McKay correspondence \cite{} describes physically stacksof D0-branes on an orbifold of $\mathbb{C}^2$ Consider a stack of D0-branes on the orbifold $\mathbb{C}^3/H$.\fdm{Why $\mathbb{C}^3$? This is not what we are doing in the rest of the paper}
The total Chan-Paton space $V_{\text{total}}$ decomposes into sectors labeled by the irreducible representations (irreps) $\rep{i}$ of the orbifold group $H$:
\begin{equation}
    V_{\text{total}} \cong \bigoplus_{i} \rep{i} \tensor \mult{i} \,,
\end{equation}
where $\mult{i} \cong \mathbb{C}^{N_i}$ denotes the multiplicity space associated with the $i$-th node of the quiver.

The fields of the original supersymmetric quantum mechanics, denoted by $\Phi$, are promoted to operators on this total space. Since $\Phi$ corresponds to a coordinate in the target space, it transforms under $H$ in a specific representation $r_\Phi$. We can decompose $\Phi$ into block components connecting the different sectors. 
The number of arrows $n_{ji}$ pointing from node $i$ to node $j$ corresponding to the field $\Phi$ is determined by the decomposition of the tensor product of the source representation and the field representation:
\begin{equation} \label{eq:SelectionRule}
    \rep{i} \otimes r_\Phi = \bigoplus_j n_{ji} \rep{j} \,.
\end{equation}
Thus, a component $\varphi_{ji}$ is non-vanishing (an arrow exists) if and only if the representation $\rep{j}$ appears in the decomposition of $\rep{i} \otimes r_\Phi$.

\subsubsection*{The Matrix Formalism}
It is extremely convenient to formalize this decomposition using matrix notation. We view the orbifold fields as block matrices acting on the vector space of representations. For the $H=\mathbb{Z}_2$ case with irreps $\rho_0$ (trivial) and $\rho_1$ (sign), any parent field $\Phi$ lifts to a $2 \times 2$ matrix:
\begin{equation}
    \mathbf{\Phi} = \begin{pmatrix} \Phi_{00} & \Phi_{01} \\ \Phi_{10} & \Phi_{11} \end{pmatrix} \,.
\end{equation}
The selection rules in Eq.~\eqref{eq:SelectionRule} translate directly into structural constraints on this matrix:
\begin{itemize}
    \item \textbf{Neutral Fields ($r_\Phi = \rho_0$):} These fields map a representation to itself ($\rho_i \to \rho_i$). Consequently, the matrix must be \textbf{diagonal}.
    \item \textbf{Charged Fields ($r_\Phi = \rho_1$):} These fields map a representation to its exchange partner ($\rho_0 \leftrightarrow \rho_1$). Consequently, the matrix must be \textbf{off-diagonal}.
\end{itemize}
We can also view this as the statement that the orbifold generator $g$ acts via conjugation by the Pauli matrix $\sigma_z$:
\begin{equation}
    \mathbf{\Phi} \xrightarrow{g} \sigma_z \mathbf{\Phi} \sigma_z^{-1} = 
    \begin{pmatrix} \Phi_{00} & 0\\ 0 & \Phi_{11} \end{pmatrix}-\begin{pmatrix} 0 & \Phi_{01} \\ \Phi_{10} & 0 \end{pmatrix}\,.
\end{equation}
This formalism allows us to translate geometric coordinates directly into quiver invariants. The generalization to more general abelian orbifolds will be used in Section \ref{sec:GeneralOrb} to construct BPS quivers of infinite families of new geometrically engineered 5d QFTs with Coulomb branches and flavor groups of arbitrarily high rank.

\subsection{Example: The \texorpdfstring{$\mathbb{C}^2/\mathbb{Z}_2 \times \mathbb{C}$}{C2/Z2 x C} Orbifold}
\label{sec:a1orbifold}
To illustrate the power of the matrix formalism, we apply it to a well-known test case: the orbifold of flat space $\mathbb{C}^3$ by $\mathbb{Z}_2$ acting on two coordinates.

We begin with the parent theory, the theory on a stack of D0-branes probing $\mathbb{C}^3$. The relevant supersymmetric quantum mechanics can be obtained via dimensional reduction to 1d of 4d, $\mathcal{N}=4$ Super Yang-Mills. %From our point of view, the resulting theory is the \emph{BPS quiver} for $\mathbb{C}^3$, described by
Its quiver has a single-node quiver with three adjoint chiral fields $X, Y, Z$ and the superpotential:
\begin{equation}
    W = \text{Tr}\left( Z [X, Y] \right) \,.
\end{equation}
The geometry corresponds to $\mathbb{C}^3$. We act with the orbifold group $H=\mathbb{Z}_2$ such that $X$ and $Y$ form the singular $\mathbb{C}^2/\mathbb{Z}_2$ fiber, while $Z$ parametrizes the smooth $\mathbb{C}$ base.
The charge table is:
\begin{equation}
\begin{array}{c|ccc}
 & Z & X & Y \\
 \hline
U(1)_{\text{gauge}} & 0 & 0 & 0 \\
H=\mathbb{Z}_2 & 0 & 1 & 1 \\
\end{array}
\end{equation}
Using our matrix dictionary, we immediately determine the field content of the daughter theory:

\begin{enumerate}
    \item \textbf{Neutral Field ($Z$):} Since $Z$ is even ($\rho_Z = \rho_0$), it maps representations to themselves. It lifts to a \emph{diagonal} matrix:
    \begin{equation}
        \hat{Z} = \begin{pmatrix} \phi_0 & 0 \\ 0 & \phi_1 \end{pmatrix} \,.
    \end{equation}
    These correspond to adjoint scalars $\phi_0, \phi_1$ on the two gauge nodes.

    \item \textbf{Charged Fields ($X, Y$):} Since $X, Y$ are odd ($\rho_{X,Y} = \rho_1$), they map representations to their exchange partners. They lift to \emph{off-diagonal} matrices:
    \begin{equation}
        \hat{X} = \begin{pmatrix} 0 & A_2 \\ A_1 & 0 \end{pmatrix} \,, \quad 
        \hat{Y} = \begin{pmatrix} 0 & B_2 \\ B_1 & 0 \end{pmatrix} \,.
    \end{equation}
    These correspond to bifundamental pairs connecting node 0 and node 1.
\end{enumerate}

To find the superpotential, we substitute these matrices into the parent expression $W = \text{Tr}(ZXY - ZYX)$:
\begin{align}
    W_{\text{Orb}} &= \text{Tr} \left[ \begin{pmatrix} \phi_0 & 0 \\ 0 & \phi_1 \end{pmatrix} \left( \begin{pmatrix} 0 & A_2 \\ A_1 & 0 \end{pmatrix} \begin{pmatrix} 0 & B_2 \\ B_1 & 0 \end{pmatrix} - \begin{pmatrix} 0 & B_2 \\ B_1 & 0 \end{pmatrix} \begin{pmatrix} 0 & A_2 \\ A_1 & 0 \end{pmatrix} \right) \right] \\
    &= \text{Tr} \left[ \begin{pmatrix} \phi_0 & 0 \\ 0 & \phi_1 \end{pmatrix} \begin{pmatrix} A_2 B_1 - B_2 A_1 & 0 \\ 0 & A_1 B_2 - B_1 A_2 \end{pmatrix} \right] \\
    &= \text{Tr} \left( \phi_0 (A_2 B_1 - B_2 A_1) \right) + \text{Tr} \left( \phi_1 (A_1 B_2 - B_1 A_2) \right) \,.
\end{align}
This is precisely the superpotential for the $\widehat{A}_1$ affine quiver gauge theory, matching the standard result for the $\mathbb{C}^2/\mathbb{Z}_2$ singularity. The matrix formalism reproduces the correct bifundamental structure and couplings automatically.

\subsubsection*{Geometry from Nilpotency}
The orbifolded BPS quiver we have just constructed contains information about the geometry of the resolved $\mathbb{C}^2/\mathbb{Z}_2$ orbifold. We will now locate the exceptional $\mathbb{P}^1$ directly as a subspace of the moduli space of the 
%(1,1)$ representation
abelian quiver with $U(1)$ gauge group at both nodes, describing the supersymmetric quantum mechanics of a D0 brane on the resolved orbifold. In original affine coordinates the orbifold singularity is located at $x=y=0$. In our matrix language, one might have naively tried to impose that the matrices $X, Y$ themselves vanish. However, this is forbidden by stability conditions, i.e. D-term constraints. However, we \emph{can} impose the next best thing: That the \emph{orbifold} invariant matrices $X^2, Y^2, X \cdot Y$ all vanish. This requires the coordinate matrices $\mathbf{X}$ and $\mathbf{Y}$ to be nilpotent. Given that they are purely off-diagonal, a priori this means that $X$ and $Y$ must be either upper triangular or lower triangular independently. But the $X\cdot Y=0$ constraint fixes both to be lower (or upper) triangular, depending on the FI-parameter choice. For a specific choice, this implies
\[  A_2 = B_2 = 0\,. \]
The surviving fields $A_1, B_1$ then parametrize the exceptional $\mathbb{P}^1$ of the $A_1$ resolution. This confirms that the locus of stable nilpotent matrices correctly identifies the resolution.

\subsection{Application: Orbifolding the Pagoda}
\label{sec:pagodamodz2}
We now apply the same construction to Reid's Pagoda. A priori we can choose from several inequivalent $\mathbb{Z}_2$-actions, however, we have to impose that the superpotential be \emph{invariant} under it. If it were simply homogeneous, this would imply that the superspace $\theta$-variables would have to transform under it, which means that $H$ wouldn't commute with the supercharges. Hence, gauging such an $H$-action would break SUSY.
Among the remaining choices, we will focus on the action defined by the following weights (see however Section \ref{sec:GeneralOrb} for quiver of more general abelian orbifolds): 
% \fdm{The charges in the second line should be 1 1 -1 -1, see the section on $\mathbb{Z}_N$. In $\mathbb{Z}_2$ $\rho=\rho^{-1}$ so we don't see it in this section.} \ac{I realize that. I put it like this because it made more sense to me for Z2, since there are only two irreps, trivial and sign. I think it will create confusion otherwise.}
\begin{equation}
\begin{array}{c|cccccc}
 & \alpha_1 & \alpha_2 & \beta_1 & \beta_2 & w_1 & w_2\\
 \hline
U(1)_{\text{gauge}} & 1 & 1 & -1 & -1 & 0 & 0 \\
H=\mathbb{Z}_2 & 0 & 0 & 1 & 1 & 1 & 1 \\
\end{array}
\end{equation}
This is only compatible with Pagodas of \emph{odd} $k$.
Using our matrix dictionary, we immediately write down the field content of the orbifolded theory:

\begin{enumerate}
    \item \textbf{Neutral Fields ($\boldsymbol{\alpha}$):} Since $r_\alpha = \rep{0}$, $\boldsymbol{\alpha}$ becomes a diagonal matrix representing arrows within each sector:
    \begin{equation}
        \boldsymbol{\alpha} \longrightarrow \hat{\boldsymbol{\alpha}} = \begin{pmatrix} \boldsymbol{\alpha}^{(00)} & 0 \\ 0 & \boldsymbol{\alpha}^{(11)} \end{pmatrix} \,.
    \end{equation}
    
    \item \textbf{Charged Fields ($\boldsymbol{\beta}$):} Since $r_\beta = \rep{1}$, $\boldsymbol{\beta}$ becomes an off-diagonal matrix connecting the two sectors:
    \begin{equation}
        \boldsymbol{\beta} \longrightarrow 
        \hat{\boldsymbol{\beta}}=
        \begin{pmatrix} 0 & \boldsymbol{\beta}^{(01)} \\ \boldsymbol{\beta}^{(10)} & 0 \end{pmatrix} \,.
    \end{equation}

    \item \textbf{Charged Adjoints ($w_i$):} Similarly, the adjoints $w_i$ ($i=1,2$) are charged and become off-diagonal:
    \begin{equation}
        w_i \longrightarrow \hat{w}_i=\begin{pmatrix} 0 & w_i^{(01)} \\ w_i^{(10)} & 0 \end{pmatrix} \,.
    \end{equation}
\end{enumerate}

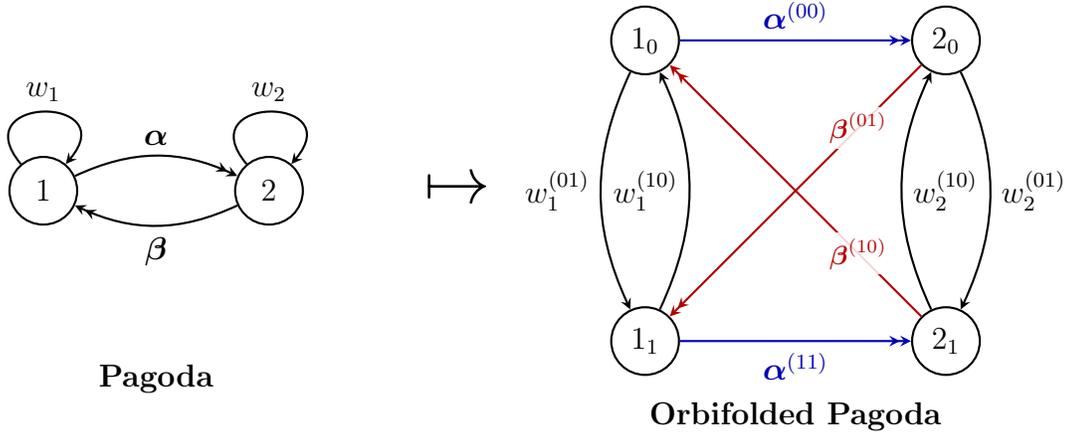
\begin{figure}[ht]
    \centering
    \begin{tikzpicture}[
        % Global Styles
        scale=1.0,
        every node/.style={transform shape},
        state/.style={circle, draw, minimum size=0.9cm, inner sep=0pt, thick, fill=white},
        % Note: \boldsymbol requires the amsmath package. 
        % If you use the 'bm' package, \bm{} is even better.
        arrowlabel/.style={font=\footnotesize, inner sep=2pt, fill=white, text opacity=1, fill opacity=0.85},
        >=stealth, thick
    ]

    % ==========================================
    % LEFT: PAGODA QUIVER
    % ==========================================
    \begin{scope}[shift={(0,0)}]
        \node[state] (L) at (0,0) {1};
        \node[state] (R) at (3,0) {2};

        % Adjoints w
        \draw[->] (L) to [out=130, in=50, loop, min distance=1.2cm] node[above] {$w_1$} (L);
        \draw[->] (R) to [out=130, in=50, loop, min distance=1.2cm] node[above] {$w_2$} (R);

        % Alpha/Beta (Bi-fundamentals) - BOLD
        \draw[->>] (L) to [bend left=25] node[above] {$\boldsymbol{\alpha}$} (R);
        \draw[->>] (R) to [bend left=25] node[below] {$\boldsymbol{\beta}$} (L);
        
        \node at (1.5, -2.5) {\textbf{Pagoda}};
    \end{scope}

    % ==========================================
    % MAPPING ARROW
    % ==========================================
    \node at (5.5, 0) {\Huge $\mapsto$};

    % ==========================================
    % RIGHT: ORBIFOLDED PAGODA QUIVER
    % ==========================================
    \begin{scope}[shift={(8,0)}]
        % -- NODES --
        \node[state] (L0) at (0, 2)   {$1_0$};
        \node[state] (R0) at (4, 2)   {$2_0$};
        \node[state] (L1) at (0, -2)  {$1_1$};
        \node[state] (R1) at (4, -2)  {$2_1$};

        % -- 1. ALPHAS (Neutral) -- BOLD
        \draw[->>, blue!70!black] (L0) -- node[above] {$\boldsymbol{\alpha}^{(00)}$} (R0);
        \draw[->>, blue!70!black] (L1) -- node[below] {$\boldsymbol{\alpha}^{(11)}$} (R1);

        % -- 2. BETAS (Charged) -- BOLD
        \draw[->>, red!70!black] (R0) -- node[arrowlabel, pos=0.25] {$\boldsymbol{\beta}^{(01)}$} (L1);
        \draw[->>, red!70!black] (R1) -- node[arrowlabel, pos=0.25] {$\boldsymbol{\beta}^{(10)}$} (L0);

        % -- 3. W FIELDS --
        \draw[->] (L0) to [bend right=25] node[midway, left] {$w_1^{(01)}$} (L1);
        \draw[->] (L1) to [bend right=25] node[midway, left] {$w_1^{(10)}$} (L0);

        \draw[->] (R0) to [bend left=25]  node[midway, right] {$w_2^{(01)}$} (R1);
        \draw[->] (R1) to [bend left=25]  node[midway, right] {$w_2^{(10)}$} (R0);
        
        \node at (2, -3.0) {\textbf{Orbifolded Pagoda}};
    \end{scope}

    % ==========================================
    % LEGEND (Centered below everything)
    % ==========================================
    % Shifted down to y = -3.5 and centered at x = 5.5
    % \node[anchor=north, align=center, font=\footnotesize, draw=black!20, thin, inner sep=5pt] 
    %     at (5.5, -3.5) {
    %     \textbf{Notation:}\\
    %     $\boldsymbol{\phi} \equiv (\phi_1, \phi_2)$ represents an $SU(2)$ flavor doublet.
    %     };

    \end{tikzpicture}
    \caption{The Quiver diagram transition. Bold symbols denote flavor doublets. Notation: $\boldsymbol{\phi} \equiv (\phi_1, \phi_2)$ represents an $SU(2)$ flavor doublet.}
    \label{fig:pagoda_orbifold}
\end{figure}

The resulting quiver is shown in Figure \ref{fig:pagoda_orbifold}.

\subsubsection{Superpotential}
The superpotential is determined by the gauge-invariant closed loops. We start with the parent superpotential:
\begin{equation}
    \label{eq:W_pagoda}
    W_{\text{Pagoda}} = \text{Tr} \Bigl( \boldsymbol{\alpha} \cdot \boldsymbol{\beta} \, w_1 \Bigr) 
    + \text{Tr} \Bigl( \boldsymbol{\beta} \cdot \boldsymbol{\alpha} \, w_2 \Bigr) 
    + \frac{1}{k+1} \text{Tr} \left( w_1^{k+1} - w_2^{k+1} \right) \,.
\end{equation}

The orbifolded superpotential is obtained by substituting the block matrices into Eq.~\eqref{eq:W_pagoda} and taking the trace.
For the polynomial term $w^{k+1}$, since $\hat{w}_i$ are off-diagonal, a non-zero trace requires an even number of insertions forming an alternating path $0 \to 1 \to 0 \dots$. The surviving terms are:
\begin{equation}
\label{eq:W_orbifold}
W_{Orb}=\tr\left(\hat{\bs\alpha}\cdot\hat{\bs\beta} \,\hat{w}_1 \right)+\tr\left(\hat{\bs\beta}\cdot\hat{\bs\alpha}\,\hat{w}_2 \right)+\frac{1}{k+1}\tr \left(\hat{w}_1^{k+1}\right)-\frac{1}{k+1}\tr \left(\hat{w}_2^{k+1}\right)
\end{equation}
whereby the trace runs over `color' and orbifold indices. Written explicitly in terms of only a trace over color indices, it would spell as follows:
\begin{equation}
    \label{eq:W_orbifold_expanded}
    \begin{aligned}
    W_{\text{Orb}} &= \text{Tr} \Bigl( \boldsymbol{\alpha}^{(00)} \cdot \boldsymbol{\beta}^{(01)} \, w_1^{(10)} \Bigr) 
                    + \text{Tr} \Bigl( \boldsymbol{\alpha}^{(11)} \cdot \boldsymbol{\beta}^{(10)} \, w_1^{(01)} \Bigr) \\
                   &+ \text{Tr} \Bigl( \boldsymbol{\beta}^{(10)} \cdot \boldsymbol{\alpha}^{(00)} \, w_2^{(01)} \Bigr) 
                    + \text{Tr} \Bigl( \boldsymbol{\beta}^{(01)} \cdot \boldsymbol{\alpha}^{(11)} \, w_2^{(10)} \Bigr) \\
                   &+ \frac{2}{k+1}  \text{Tr} \Bigl( \underbrace{w_1^{(0, 1)} w_1^{(1, 0)} \dots w_1^{(1, 0)}}_{k+1 \text{ terms}} \Bigr) \\
                   &- \frac{2}{k+1} \text{Tr} \Bigl( \underbrace{w_2^{(0, 1)} w_2^{(1, 0)} \dots w_2^{(1, 0)}}_{k+1 \text{ terms}} \Bigr) \,.
    \end{aligned}
\end{equation}

% ==========================================
% SECTION 4: GEOMETRY OF THE MODULI SPACE
% ==========================================
\section{Geometry of the Moduli Space}\label{sec:OrbGeom}
The geometry of the threefold is defined by the vacuum moduli space of the theory. This corresponds to the set of holomorphic gauge-invariant operators (mesons) subject to the constraints imposed by the F-term equations.

\subsection{The Determinantal Variety}
We define the orbifold geometry by first determining the transformation properties of the parent coordinates $u,v,z,w$ under the $H=\mathbb{Z}_2$ action. 
Recall the charge assignments from the orbifold construction:
\begin{itemize}
    \item The base fields $\alpha$ are neutral.
    \item The fiber fields $\beta$ and the adjoints $w$ are charged.
\end{itemize}
Consequently, as it is clear from \eqref{eq:pagodacordstoquiver}, all four affine coordinates of the parent Pagoda theory are odd under the orbifold action:
\begin{equation}
    (u,v,w,z) \quad \longrightarrow \quad (-u,-v,-w,-z). 
\end{equation}
% \begin{equation}
%     \begin{aligned}
%         u &\sim \alpha_1 \beta_2 \longrightarrow (+1)(-1) = -1 \,, \\
%         v &\sim \alpha_2 \beta_1 \longrightarrow (+1)(-1) = -1 \,, \\
%         z &\sim \alpha_1 \beta_1 \longrightarrow (+1)(-1) = -1 \,, \\
%         w &\longrightarrow -1 \,.
%     \end{aligned}
% \end{equation}

Since the coordinates flip sign, %there are no linear gauge-invariant operators. T
the ring of invariants is generated by all the monomials of degree two in $u,v,z,w$. There are ten such generators, which we arrange into a symmetric $4 \times 4$ matrix $\mathcal{M}$: 
% \fdm{Should this be an $\otimes$ instead of a $\cdot$?} \ac{No, these are ordinary products. It's the outer product of two vectors, but the entries are ordinary complex numbers.}
\begin{equation} \label{eq:determinantal}
    \mathcal{M} = \begin{pmatrix} u\\v\\z\\w \end{pmatrix} \cdot 
    \begin{pmatrix} u & v & z & w \end{pmatrix}
    = \begin{pmatrix} 
    u^2 & uv & uz & uw \\ 
    vu & v^2 & vz & vw \\ 
    zu & zv & z^2 & zw \\ 
    wu & wv & wz & w^2 
    \end{pmatrix} \,.
\end{equation}
The geometry is defined by two conditions on this matrix:
\begin{enumerate}
    \item \textbf{Rank Condition:} Since $\mathcal{M}$ is constructed from the outer product of the vector $(u,v,z,w)$, it must have rank one. This implies the vanishing of all $2 \times 2$ minors, which are the defining equations of the orbifold singularity $\mathbb{C}^4 / \mathbb{Z}_2$.
    \item \textbf{Dynamical Constraint:} The parent Pagoda equation $uv - z^2 + w^{2k} = 0$ imposes the following constraint on the matrix entries:
    \begin{equation}
        \mathcal{M}_{12} - \mathcal{M}_{33} + (\mathcal{M}_{44})^k = 0 \,.
    \end{equation}
\end{enumerate}
The Orbifolded Pagoda is the intersection of these two conditions and it is a determinantal variety.

\subsection{Resolution and the Hidden Surface}\label{Sec:ResoHiddenSurface}
We can determine the topology of the compact 4-cycles by resolving the orbifold singularity. Our threefold will be defined as a hypersurface in the ambient space $\mathbb{C}^4/\mathbb{Z}_2$, which is resolved by a single blowup introducing an exceptional divisor $E \cong \mathbb{P}^3$. In toric language, this resolution is described by a Gauge Linear Sigma Model (GLSM) with a $U(1)$ gauge field corresponding to the blow-up parameter $\lambda$.\footnote{Note that the GLSM is describing the toric variety resolving $\mathbb{C}^4/\mathbb{Z}_2$, while our nontoric threefold is described by an algebraic equation in this space.} The charge assignment is:
\begin{equation}\label{Eq:PesiC4modZ2}
\begin{array}{c|ccccc}
 & u & v & z & w & \lambda \\
\hline
\mathbb{C}^* & 1 & 1 & 1 & 1 & -2
\end{array}
\end{equation}
The D-term constraint implies that $u,v,z,w$ cannot simultaneously vanish, defining a $\mathbb{P}^3$ base, while $\lambda$ is a coordinate on the fiber of the line bundle $\mathcal{O}(-2)$. The exceptional divisor is located at $\lambda=0$.

The geometry of the Calabi-Yau is the \emph{proper transform} of the orbifolded hypersurface. Under the blowup with scaling parameter $\lambda$, the defining equation of homogeneous degree two is given by
\begin{equation}\label{Eq:orbifoldresolved}
    uv - z^2 +  w^{2k}\,\lambda^{k-1} = 0\,.
\end{equation}
The exceptional divisor corresponds to the locus $\lambda=0$:
\begin{itemize}
    \item For $k=1$, the equation is $uv - z^2 + w^2 = 0$. This defines a smooth quadric surface in $\mathbb{P}^3$, which is isomorphic to $\mathbb{P}^1 \times \mathbb{P}^1 \cong \mathbb{F}_0$.
    \item For $k > 1$, the higher-order term vanishes, leaving the equation $uv - z^2 = 0$. This defines a singular quadric cone in $\mathbb{P}^3$. The minimal resolution of this cone yields the Hirzebruch surface $\mathbb{F}_2$.
\end{itemize}
Thus, the geometry of the orbifolded Pagoda contains a compact divisor of type $\mathbb{F}_2$.

To see why the singular quadric in $\mathbb{P}^3$ corresponds to a partially blown-down $\mathbb{F}_2$, we describe the latter torically as $\mathbb{P}\left(\mathcal{O}_{\mathbb{P}^1}(0)\oplus \mathcal{O}_{\mathbb{P}^1}(-2)\right)$, which is the same as the following toric variety:
\begin{equation}\label{Eq:F2GLSM}
\mathbb{F}_2:\qquad \begin{array}{c|cccc}
 & x_1 & x_2 & y_1 & y_2 \\
\hline
\mathbb{C}^*_1 & 1 & 1 & 0 & -2\\
\mathbb{C}^*_2 & 0 & 0 & 1 & 1
\end{array}
\end{equation}
We can blow down the K\"ahler modulus associated to the first $\mathbb{C}^*$-action. This is achieved by writing its invariants. The resulting space is given by the ambient $\mathbb{P}^3$:
\begin{equation}
\begin{array}{c|ccccc}
 & u:=x_1^2 y_2 & v:=x_2^2 y_2 & z:=x_1 x_2 y_2 & w:=y_1  \\
\hline
\mathbb{C}^*_2 & 1 & 1 & 1 & 1  
\end{array}
\end{equation}
subject to the relation $u\,v = z^2$. So, this describes an $\mathbb{F}_2$ with its $(-2)$-curve blown down to an $A_1$-singularity. This singular surface will be a central character to our story.

The appearance of the compact surface $\mathbb{F}_2$ already indicates the existence of a rank-one Coulomb branch and, in an appropriate K\"ahler chamber, a low-energy gauge theory phase. In particular, when the exceptional divisor is a genuine $\mathbb{F}_2$, there will be an $SU(2)$ gauge theory phase where the Coulomb parameter will be described by the fiber class, while the volume of the base $(-2)$-curve plays the role of the inverse bare coupling. We now turn to the analysis of this curve.

\subsection{The Pagodina Curve} \label{sec:pagodinaaffine}
The most critical physics lies in the \emph{singularity} of this divisor itself, which is also a singularity of the threefold.
To see the local geometry experienced by the BPS states, we analyze the local neighborhood of the ``north pole'' of the exceptional $\mathbb{P}^3$ (where $w \neq 0$). In the affine patch where we set $w=1$, the proper transform equation becomes:
\begin{equation}
    \label{eq:pagoda_recursion}
    uv - z^2 + \lambda^{k-1} = 0 \,.
\end{equation}
Since we are focusing on odd values of $k$, this is precisely the defining equation of a Reid Pagoda of order $p=(k-1)/(2)$. We will affectionately refer to the associated curve as a \emph{Pagodina}\footnote{Given the composition of the author list, the use of the Italian diminutive was statistically inevitable.} curve. It will be our second main character, as it will host unconventional 5d matter.

This yields a striking structural observation: The $\mathbb{Z}_2$ orbifold of a Pagoda of order $k$ contains, within its resolution, a local geometry equivalent to a Pagoda of order $(k-1)/2$.

Resolving the Pagodina singularity blows up a $\mathbb{CP}^1$, the Pagodina curve, and turns the exceptional divisor into a smooth Hirzebruch surface $\mathbb{F}_2$. It is straightforward to see that the Pagodina curve is precisely the rigid $(-2)$-curve of $\mathbb{F}_2$ (given by $y_2=0$ in \eqref{Eq:F2GLSM}). We now do it explicitly.

Let us start from the equation \eqref{Eq:orbifoldresolved} in the toric space \eqref{Eq:PesiC4modZ2} and perform the shift
\begin{equation}\label{Eq:Singular3foldWithPagodina}
    z = t + w^{k}\,\lambda^{(k-1)/2}\,.
\end{equation}
With this redefinition, the hypersurface equation becomes
\begin{equation}\label{Eq:PagodinaThreefoldGlobalWitht}
    uv = t\left(t + 2 w^{k}\,\lambda^{(k-1)/2}\right)\,.
\end{equation}
The singular locus is located at $u=v=t=\lambda=0$. 
We can resolve this space by blowing up the ideal $(u,t)$. The resulting space is given by the equation 
\begin{equation}\label{Eq:PagodinaResolution}
    uv = t\,\left(\sigma\,t + 2 w^{k}\,\lambda^{(k-1)/2} \right)= 0\,
\end{equation}
in the toric space 
\begin{equation}\label{Eq:PesiC4modZ2Resolved}
\begin{array}{c|cccccc}
 & u & v & t & w & \lambda & \sigma \\
\hline
\mathbb{C}^* & 1 & 1 & 1 & 1 & -2 & 0 \\
\mathbb{C}^* & 1 & 0 & 1 & 0 & 0 & -1 \\
\end{array}
\end{equation}
The divisor $\lambda=0$ in the resolved space is a Hirzebruch surface $\mathbb{F}_2$. The map between this presentation and the standard toric description of $\mathbb{F}_2$ (see \eqref{Eq:F2GLSM}) is given by
\begin{equation}
    u = x_1^2, \qquad 
    v = x_2^2 y_2, \qquad 
    t = x_1 x_2, \qquad 
    \sigma = y_2\,.
\end{equation}
Under this identification, the Pagodina curve corresponds to the locus $y_2=0$ in the standard toric presentation, which in the present coordinates is given by $v=\sigma=0$.

Instead of resolving the singularity, one may consider deforming it.
Working in the patch $w=1$, where the singularity is located, the geometry is
described by the Reid pagoda hypersurface
\begin{equation}
\label{eq:pagoda_simple}
    uv = z^2 - \lambda^{2p},
\end{equation}
with $p=\frac{k-1}{2}$.
This singularity admits $2p-1$ versal complex structure deformations.
However, only $p$ of them are 5d dynamical degrees of freedom in M-theory in the sense of
Refs.~\cite{Gukov:1999ya,Shapere:1999xr},\footnote{Let us review the counting of the dynamical modes: the singularity \eqref{eq:pagoda_simple} is quasi-homogeneous. Normalizing its quasi-homogeneous weights $q_u, q_v, q_z, q_{\lambda}$ so that the polynomial in \eqref{eq:pagoda_simple} has weight one, we have 
$ %\begin{equation*}
    q_{u} = q_{v}= q_{z}= \frac{1}{2}, \quad q_{\lambda} = \frac{1}{2p}. 
$ % \end{equation*}
Now, define 
$ %\begin{equation*}
    \hat{c} = 4 - 2(q_u + q_v + q_z + q_{\lambda}) = \frac{p-1}{p},
$ %\end{equation*}
and $Q_{j} \equiv j q_{\lambda}$. 
Following \cite{Gukov:1999ya,Shapere:1999xr}, if 
$Q_{j}  \leq \frac{\hat{c}}{2} $ (or equivalently $j \leq p-1$),
%    $Q_{j}  < \frac{\hat{c}}{2} $ (or equivalently $j \leq p-2$),
the deformation is 
dynamical in M-theory. Recently, it was argued that the dynamical deformations must satisfy instead the strict inequality $Q_{j}  < \frac{\hat{c}}{2} $ \cite{Acharya:2024bnt}. However, this does not alter our result in any significant way.   
} namely
\begin{equation}
\label{eq:generic_def}
    uv = z^2 - \lambda^{2p}
    + \sum_{i=0}^{p-1} c_i \lambda^i \: .
\end{equation}

In the five-dimensional theory obtained from M-theory on this threefold,
the dynamical deformations correspond to giving vacuum expectation values
to dynamical hypermultiplets.
The fact that M-theory on the Reid pagoda exhibits a $p$-dimensional Higgs
branch is well established in the literature
\cite{Closset:2020scj,Sangiovanni:2024nfz}.\footnote{Notice that in the gauge theory regime, the Pagodina curve is blown up and the corresponding Pagoda matter become massive.}

\

Let us now return to the full threefold~\eqref{Eq:PagodinaThreefoldGlobalWitht}, restoring the coordinate
$w$.
In this case, the dynamical deformations take the form
\begin{equation}
\label{Eq:normalDefPagodina}
    uv = t\left(t + 2 w^{k}\lambda^{(k-1)/2}\right)
    + \sum_{i=0}^{(k-3)/2} c_i \, w^{2i+2}\lambda^i \: .
\end{equation}
%The deformation~\eqref{Eq:normalDefPagodina} completely smooths the singularity.

Let us consider the $k=3$ case first (meaning $p=1$). The only dynamical deformation is the monomial $w^2$. The equation \eqref{Eq:normalDefPagodina} becomes 
\begin{equation}
\label{Eq:normalDefPagodina2}
    uv = t\left(t + 2 w^{3}\lambda\right)
    +  c_0 \, w^{2} \: .
\end{equation}
The exceptional divisor at $\lambda = 0$ is now a smooth quadric in $\mathbb{P}^3$, namely
$\mathbb{F}_0 = \mathbb{P}^1 \times \mathbb{P}^1$.

Although the defining equation is formally similar to the deformation of the local $\mathbb{F}_2$ geometry to the local $\mathbb{F}_0$ considered by \cite{Jefferson:2018irk},  the physical interpretation is different. 
In the construction of \cite{Jefferson:2018irk}, the deformation corresponds to a non-normalizable complex structure deformation, which geometrically implements a Hanany-Witten transition. 
In contrast, in our case the presence of the higher-degree term $w^6\lambda^2$ (equivalently $t\,w^3\lambda$ in \eqref{Eq:normalDefPagodina}) renders the deformation normalizable in the sense of~\cite{Gukov:1999ya,Shapere:1999xr}. 
%As a result, the deformation becomes dynamical in five dimensions and cannot be interpreted as a mere background deformation or as a Hanany--Witten move. 
This distinction is crucial: while the \cite{Jefferson:2018irk} deformation encodes a non-dynamical rearrangement of branes, our geometry describes a genuinely dynamical deformation of the five-dimensional theory.

An even more surprising phenomenon occurs for $k\geq 5$.\footnote{In this case, the Pagodina singularity is a genuine Reid Pagoda, while for $k=3$ the pointlike singularity on the $w=1$ patch is a conifold.}
In this case, one can deform the Pagoda singularity of the threefold by switching on all deformations proportional to $\lambda$, while keeping only the $w^2$ deformation turned off. E.g. for $k=5$ we can deform the Pagodina singularity as 
\begin{equation}
\label{Eq:normalDefPagodina3}
    uv = t\left(t + 2 w^{5}\lambda^2\right)
    +  c_1 \, w^{4}\lambda \: .
\end{equation}
The resulting threefold is completely smooth.\footnote{This is manifest from the fact that the deformed hypersurface \eqref{Eq:normalDefPagodina} no longer admits
a factorized form that would allow for a small resolution.}
As a consequence, there is no remaining singular locus that could support a
crepant resolution, and hence no exceptional $\mathbb{P}^1$ (notice that the exceptional divisor at $\lambda=0$ is still singular). In other words, the K\"ahler modulus is frozen. Physically this corresponds to \emph{freezing the theory at infinite coupling}.

%Thus, while the singularity admits a small resolution at the undeformed point,turning on generic normalizable deformations removes the possibility of anycrepant resolution.
Summing up, while the singularity admits a small resolution at the undeformed point,
generic normalizable deformations proportional to $\lambda$ eliminate any
crepant resolution, effectively freezing a K\"ahler modulus at infinite gauge
coupling. The quadratic $w^2$ deformation is an exception, as it does not lead to such a freezing.

\subsection{The Matrix Dictionary}
To rigorously identify the geometric loci, we translate the affine coordinates $u,v,w,z$ of the parent theory  into the operator language of the orbifolded quiver.
Under the orbifold map, the scalar coordinates lift to $2 \times 2$ matrices.

Using the matrix forms of the fields established in Section 3.2, we define the Coordinate Matrices.
The base coordinates $\hat{\mathrm{U}}, \hat{\mathrm{V}}, \hat{\mathrm{Z}}$ are defined exactly as in the parent theory:
\begin{equation}
    \hat{\mathrm{U}} \equiv 2\, \hat{\alpha}_1 \, \hat{\beta}_1 \,, \quad
    \hat{\mathrm{V}} \equiv 2 \,\hat{\alpha}_2 \, \hat{\beta}_2 \,, \quad
    \hat{\mathrm{Z}} \equiv \hat{\alpha}_1 \, \hat{\beta}_2 + \hat{\alpha}_2 \, \hat{\beta}_1 \,.
\end{equation}
For the deformation coordinate, the quiver contains two adjoints $w_1, w_2$, we define two matrices $\hat{\mathrm{W}}_i$:
\begin{equation}
    \hat{\mathrm{W}}_i = \begin{pmatrix} 0 & w^{(01)}_i \\ w^{(10)}_i & 0 \end{pmatrix} \,.
\end{equation}
Note that all the matrices $\hat{\mathrm{U}}, \hat{\mathrm{V}}, \hat{\mathrm{Z}}, \hat{\mathrm{W}}_i$ transform in the sign representation of $\mathbb{Z}_2$ (odd parity), consistent with their off-diagonal structure.

\subsection{Geometric loci via nilpotent quiver representations}
We can now identify the exceptional geometry by exploiting via quiver quantum mechanics the matrix structure of the orbifold coordinates, much like we did in the basic example of $\mathbb{C}^2/\mathbb{Z}_2\times \mathbb C$, in section \ref{sec:a1orbifold}.
The singular point of the affine variety is the locus where all coordinate invariants vanish: $u=v=z=w=0$. 
In our matrix formalism, this translates to the condition that all coordinate matrices $\hat{\mathrm{U}}, \hat{\mathrm{V}}, \hat{\mathrm{Z}}, \hat{\mathrm{W}}_i$ are nilpotent. Stability conditions prevent us from setting these matrices to zero, however we \emph{can} set all orbifold invariants to zero. These invariants are constructed analogously to \eqref{eq:determinantal}, by defining the following \emph{non-Abelian field valued matrix}:
\begin{equation} \label{eq:matrix_determinantal}
    \setlength{\arraycolsep}{3pt}
    \boldsymbol{\mathcal{M}} = \begin{pmatrix} \hat{\mathrm{U}} \\ \hat{\mathrm{V}} \\ \hat{\mathrm{Z}} \\ \hat{\mathrm{W}}_i \end{pmatrix} \cdot 
    \begin{pmatrix} \hat{\mathrm{U}} & \hat{\mathrm{V}} & \hat{\mathrm{Z}} & \hat{\mathrm{W}}_j \end{pmatrix}
    = \begin{pmatrix} 
    \hat{\mathrm{U}}^2 & \hat{\mathrm{U}} \cdot \hat{\mathrm{V}} & \hat{\mathrm{U}} \cdot \hat{\mathrm{Z}} & \hat{\mathrm{U}} \cdot \hat{\mathrm{W}}_j \\ 
    \hat{\mathrm{V}} \cdot \hat{\mathrm{U}} & \hat{\mathrm{V}}^2 & \hat{\mathrm{V}} \cdot \hat{\mathrm{Z}} & \hat{\mathrm{V}} \cdot \hat{\mathrm{W}}_j \\ 
    \hat{\mathrm{Z}} \cdot \hat{\mathrm{U}} & \hat{\mathrm{Z}} \cdot \hat{\mathrm{V}} & \hat{\mathrm{Z}}^2 & \hat{\mathrm{Z}} \cdot \hat{\mathrm{W}}_j \\ 
    \hat{\mathrm{W}}_i \cdot \hat{\mathrm{U}} & \hat{\mathrm{W}}_i \cdot \hat{\mathrm{V}} & \hat{\mathrm{W}}_i \cdot \hat{\mathrm{Z}} & \hat{\mathrm{W}}_i\hat{\mathrm{W}}_j 
    \end{pmatrix} \,.
\end{equation}
Now, to find the exceptional locus of the resolution, we simply set $\boldsymbol{\mathcal{M}}=0$.
Crucially, nilpotency does \emph{not} imply that the matrices themselves vanish.
In a resolved phase, the D-term stability conditions (FI parameters) prevent the fields from vanishing identically.
\begin{equation}
    \sum |Q_{\text{in}}|^2 - \sum |Q_{\text{out}}|^2 = \theta > 0 \,.
\end{equation}
For our quiver quantum mechanics, we will choose the following FI parameters:
\begin{equation}
\label{eq:orderingtheta}
    \theta_{L0}<\theta_{L1}<\theta_{R0}<\theta_{R1}
\end{equation}
subject to $\theta_{L0}+\theta_{L1}+\theta_{R0}+\theta_{R1}=0$.

Seeing that all  matrices $\hat{\mathrm{U}}, \hat{\mathrm{V}}, \hat{\mathrm{Z}}, \hat{\mathrm{W}}_i$ are of the form
\[\begin{pmatrix}
0&*\\ * &0   
\end{pmatrix}
\]
the diagonal of $\boldsymbol{\mathcal{M}}$ tells us that all four must be nilpotent. Then, the fact that their products are all zero tells us that they are all simultaneously either upper or lower triangular. Our stability choice is such that they must all be upper triangular. We can now explore the compact cycles of our geometry.

\subsubsection*{The $\mathbb{F}_2$ Surface}

To find the $\mathbb{F}_2$ divisor, we search for a configuration where the coordinate matrices are nilpotent but non-zero.
Repeating the matrices here for convenience:
\begin{equation} \label{eq:matrix_definitions}
\begin{aligned}
    \hat{\mathrm{U}} &= 2 \begin{pmatrix} 0 & \alpha_1^{(00)} \beta_1^{(01)} \\ \alpha_1^{(11)} \beta_1^{(10)} & 0 \end{pmatrix} \,, \quad
    &\hat{\mathrm{V}} &= 2 \begin{pmatrix} 0 & \alpha_2^{(00)} \beta_2^{(01)} \\ \alpha_2^{(11)} \beta_2^{(10)} & 0 \end{pmatrix} \,, \\[10pt]
    \hat{\mathrm{Z}} &= \begin{pmatrix} 0 & \alpha_1^{(00)} \beta_2^{(01)} + \alpha_2^{(00)} \beta_1^{(01)} \\ \alpha_1^{(11)} \beta_2^{(10)} + \alpha_2^{(11)} \beta_1^{(10)} & 0 \end{pmatrix} \,, \quad
    &\hat{\mathrm{W}}_i &= \begin{pmatrix} 0 & w^{(01)}_i \\ w^{(10)}_i & 0 \end{pmatrix} \,.
\end{aligned}
\end{equation}
% \mdm{metterei solo “it is immediate to check that
% the minimal condition to impose is \begin{equation}\label{Eq:F2inQuiverEqn}
%     \mathbb{F}_2: \qquad \boldsymbol{\beta}^{(10)} \equiv \begin{pmatrix} \beta_1^{(10)} \\ \beta_2^{(10)} \end{pmatrix} = 0 \quad \text{and} \quad \boldsymbol{w}^{(10)}\equiv \begin{pmatrix} w_1^{(10)} \\ w_2^{(10)} \end{pmatrix}=0\,.
% \end{equation}...”}
We see that the minimal constraints we must impose to make them all upper triangular are achieved by setting the "return" arrows to zero, i.e. 
%Note that since $\boldsymbol{\beta}$ is an $SU(2)$ doublet, this condition requires the vanishing of the entire vector block:
\begin{equation}\label{Eq:F2inQuiverEqn}
    \mathbb{F}_2: \qquad \boldsymbol{\beta}^{(10)} \equiv \begin{pmatrix} \beta_1^{(10)} \\ \beta_2^{(10)} \end{pmatrix} = 0 \quad \text{and} \quad \boldsymbol{w}^{(10)}\equiv \begin{pmatrix} w_1^{(10)} \\ w_2^{(10)} \end{pmatrix}=0\,.
\end{equation}
This forces all closed loops to vanish (nilpotency). The stability condition $\xi > 0$ ensures that we are on the resolved branch of the vacuum moduli space, meaning the fields parametrizing the cycle cannot all vanish simultaneously. The surviving fields $\boldsymbol{\alpha}^{(00)}$, $\boldsymbol{\alpha}^{(11)}$, $\boldsymbol{\beta}^{(01)}$ and $\boldsymbol{w}^{(01)}$ serve as homogeneous coordinates on the exceptional divisor.

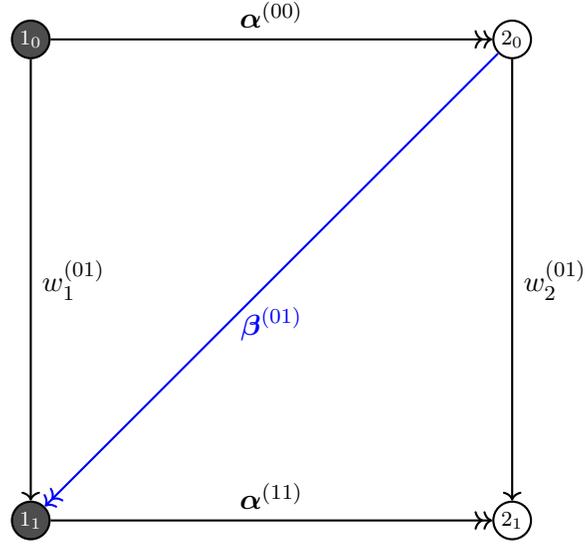
\begin{figure}
\begin{center}
$%\begin{equation}
\begin{tikzpicture}[
  x=1.6cm,y=1.6cm,
  place/.style={circle,draw=black!500,fill=black!70,thick,inner sep=0pt,minimum size=5mm}
]
% Nodi (stile CODE 1)
\node at (-2, 0) [place,fill=black!70] (N1) {};
\node at ( 2, 0) [place,fill=white]    (N2) {};
\node at ( 2,-4) [place,fill=white] (N3) {};
\node at (-2,-4) [place,fill=black!70]    (N4) {};

% Numeri dei nodi (contrasto con il fill)
\node[white] at (-2, 0) {\scriptsize $1_0$};
\node[black] at ( 2, 0) {\scriptsize $2_0$};
\node[black] at ( 2,-4) {\scriptsize $2_1$};
\node[white] at (-2,-4) {\scriptsize $1_1$};

% Lati del perimetro (->> come nell'originale)
\draw[thick,->] (N1) to[bend left=0] node[midway,right] {\small $w_{1}^{(01)}$} (N4);
\draw[thick,->] (N2) to[bend left=0] node[midway,right] {\small $w_{2}^{(01)}$} (N3);
\draw[thick,<<-] (N3) to[bend left=0] node[midway,above] {\small $\boldsymbol{\alpha}^{(11)}$} (N4);
\draw[thick,->>] (N1) to[bend left=0] node[midway,above]  {\small $\boldsymbol{\alpha}^{(00)}$} (N2);

% Diagonali (etichette spostate: più "above" e più "below")
%\draw[thick,->,red] (N1) to[bend right=0]
%node[midway,above,yshift=12pt,xshift=-1pt] {\small $w_{1}^{(01)}$} (N3);
\draw[thick,->>,blue] (N2) to[bend right=0]
  node[midway,below,yshift=-6pt] {\small $\boldsymbol{\beta}^{(01)}$} (N4);
\end{tikzpicture}\nonumber
$%\end{equation}
\end{center}
    \caption{Subquiver with $\mathbb{F}_2$ as its moduli space.}
    \label{eq:quiverlocalf2sliced}
\end{figure}

It is easy to see that, upon restricting to this locus, we recover the quiver (see Figure~\ref{eq:quiverlocalf2sliced}) and F-term relations of the NCCR of the $\mathbb F_2$ surface \cite{craw2008projective}. 
The GLSM related to the quiver in Figure~\ref{eq:quiverlocalf2sliced} describes the $\mathbb{F}_2$ surface as the moduli space of the the abelian quiver with all $U(1)$ gauge groups\footnote{Equivalently, the quiver representation with dimension vector $\Vec{d}=(1,1,1,1)$.}. It is given by
\begin{equation}
 \label{eq:GLSMF2quiver}
    \begin{array}{c|cccccccc|c}
       & \alpha_1^{(00)} & \alpha_2^{(00)} & \alpha_1^{(11)} & \alpha_2^{(11)} & \beta_1^{(01)} & \beta_2^{(01)} &
       w_1^{(01)}  &
        w_2^{(01)}  & \text{FI}
       \\
       \hline
      \mathbb C^*_1  & 1 & 1  & 0&0&0&0&1&0& \xi_1  \\
      \mathbb C^*_2  & 0 & 0  & 1&1&0&0&0&1& \xi_2  \\
      \mathbb C^*_3  & 1 & 1  &1&1&-1&-1&0&0& \xi_3  \\
       \mathbb C^*_4  & 0 & 0  & 0&0&1&1&1&1& \xi_4  \\
    \end{array}
\end{equation}
where in the last column we have written the corresponding combinations of the stability parameters that are positive in the chamber we chose (i.e. $\theta_{L0}<\theta_{L1}<\theta_{R0}<\theta_{R1}$): 
\begin{equation}
    \xi_1= -\theta_{L0}\:,\quad \xi_2= \theta_{R1}\:, \quad\xi_3=\theta_{R0}+\theta_{R1}=-\theta_{L0}-\theta_{L1}\:, \quad
    \xi_4=\theta_{L1}+\theta_{R1} 
\end{equation}
The relations that survive after imposing \eqref{Eq:F2inQuiverEqn} are
\begin{equation}\label{Eq:relationsF2GLSM}
    A \begin{pmatrix}
        \beta_1^{(01)} \\ \beta_2^{(01)}
    \end{pmatrix} =0 \quad\mbox{ and }\quad 
    \begin{pmatrix}
        w_2^{(01)} & w_1^{(01)}
    \end{pmatrix} A =0 \qquad\mbox{with }\quad A=\begin{pmatrix}
        \alpha_1^{(00)} & -\alpha_2^{(00)} \\
        -\alpha_1^{(11)} & \alpha_2^{(11)} \\
    \end{pmatrix}\:.
\end{equation}

\subsubsection*{The Pagodina via Vertical Quiver Fusion}
Having established that the threefold contains a  $\mathbb{F}_2$ divisor, and having isolated it in the quiver moduli space language, we will now show in quiver language that there is a patch in the moduli space where the threefold looks like an order $(k-1)/2$ Pagoda. 

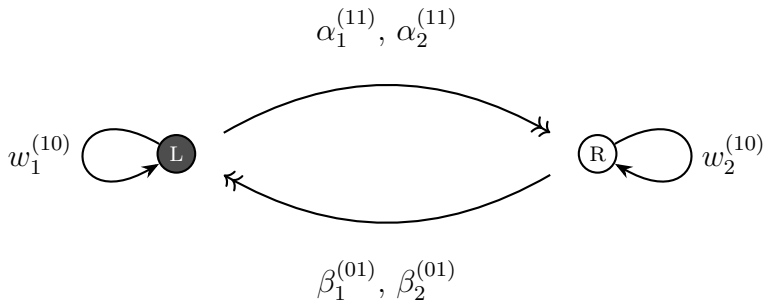
\begin{figure}
\begin{center}
\begin{tikzpicture}[
  x=1.4cm,y=1.4cm,
  place/.style={circle,draw=black!500,fill=black!70,thick,inner sep=0pt,minimum size=5mm}
]
% --- Nodi (stile CODE 1) ---
\node at (-2,0) [place,fill=black!70] (L) {};
\node at (  2,0) [place,fill=white] (R) {};

% Etichette dei nodi (contrasto con il fill)
\node[white] at (-2,0) {\scriptsize L};
\node[black] at ( 2,0) {\scriptsize R};

% Loop su ciascun nodo 
\draw[thick,-Stealth,looseness=15,out=150,in=210] (L) to node[left]  {${w}_{1}^{(10)}$} (L);
\draw[thick,-Stealth,looseness=15,out=30,in=-30]  (R) to node[right] {${w}_{2}^{(10)}$} (R);

\draw[thick,->>]  (-1.55,0.20) [bend left]  to (1.55,0.20);
\node at (0,1.2) {${\alpha}_{1}^{(11)},\,{\alpha}_{2}^{(11)}$};

\draw[thick,<<-] (-1.55,-0.20) [bend right] to (1.55,-0.20);
\node at (0,-1.2) {${\beta}_{1}^{(01)},\,{\beta}_{2}^{(01)}$};
\end{tikzpicture}
\end{center}
\caption{Quiver after Higgsing.}
\label{fig:quiverlocalf2higgsed}
\end{figure}

The Pagodina curve is obtained by setting $\boldsymbol{\beta}^{(01)}=0$ in the space defined by
\eqref{eq:GLSMF2quiver} and \eqref{Eq:relationsF2GLSM}. From the last row of
\eqref{eq:GLSMF2quiver}, we see that $\boldsymbol{w}^{(01)} \neq (0,0)$. Together with the relations
\eqref{Eq:relationsF2GLSM} and the stability conditions $\xi_1>0$ and $\xi_2>0$, this implies that both $w_1^{(01)}\neq 0$ and $w_2^{(01)}\neq 0$.
We may therefore gauge-fix them to $1$ in a neighborhood of the curve. 
This gauge-fixing identifies pairs of nodes in the original quiver, yielding the quiver shown in
Figure~\ref{fig:quiverlocalf2higgsed}. After this identification, the GLSM describing the threefold
in this local region takes the form
\begin{equation}
 \label{eq:GLSMpagodaz2ClosetoC1}
    \begin{array}{c|cccccccccc|c}
       & \alpha_1^{(00)} & \alpha_2^{(00)} & \alpha_1^{(11)} & \alpha_2^{(11)} & \beta_1^{(01)} & \beta_2^{(01)} & \beta_1^{(10)} & \beta_2^{(10)} &
       w_1^{(10)}  &
        w_2^{(10)}  & \text{FI}
       \\
       \hline
      \mathbb C^* & 1 & 1 & 1 & 1 & -1 & -1 & -1 & -1 & 0 & 0 &  \xi_3>0  \\
    \end{array}
\end{equation}
The superpotential is obtained by  inserting $w_{1}^{(01)} = w_{2}^{(01)} = 1$ into \eqref{eq:W_orbifold_expanded}. This produces several quadratic terms, that allow us to eliminate some variables; in particular they force
${\beta}_{i}^{(10)}=\tfrac12{\beta}_{i}^{(01)}(w_{1}^{(10)}-w_{2}^{(10)})$ and ${\alpha}_{i}^{(00)}={\alpha}_{i}^{(11)}$. %\equiv {\alpha}_{i}$. 
The resulting effective superpotential governing the remaining degrees of freedom is
{\small\begin{equation}
    W_{\text{eff}} = \text{Tr} \Bigl( \boldsymbol{\alpha}^{(11)} \cdot \boldsymbol{\beta}^{(01)} \, w_1^{(10)} \Bigr) 
    + \text{Tr} \Bigl( \boldsymbol{\beta}^{(01)} \cdot \boldsymbol{\alpha}^{(11)} \, w_2^{(10)} \Bigr) 
    + \frac{1}{p}\text{Tr}\left(\left({w}_1^{(10)}\right)^{p+1} - \left({w}_2^{(10)}\right)^{p+1}\right),\nonumber
\end{equation}}
with the corresponding maps represented in the quiver of
Figure~\ref{fig:quiverlocalf2higgsed}. These are precisely the quiver and superpotential of the
Reid pagoda of order  $p = \tfrac{k-1}{2}$.

The geometry in a neighborhood of the Pagodina curve is therefore locally isomorphic to the
Reid pagoda threefold. Shrinking this curve produces a point-like Pagoda singularity localized on
the exceptional divisor. This sharply contrasts with the local $\mathbb{F}_2$
threefold, where collapsing the (-2)curve gives rise instead to a \emph{line} of $A_1$
singularities.

We can identify the representation corresponding to a D2-brane wrapping this Pagodina curve. It would correspond to the Higgsed quiver \Cref{fig:quiverlocalf2higgsed} with $U(0) \times U(1)$ gauge group. In terms of quiver representations, this is the $(0,1)$ representation,
% This corresponds to the $(0,1)$-representation (i.e. with $U(0) \times U(1)$ gauge group). 
and it has a zero-dimensional moduli space corresponding to a fat point $\mathbb{C}[w_2^{(10)}]/\left(\left(w_2^{(10)}\right)^p\right)$. We can easily find the lift of this representation to the full orbifold quiver \Cref{fig:pagoda_orbifold}: the $2_0$ and $2_1$ nodes have $U(1)$ gauge group each, and the other two are empty, i.e. the representation with dimension vector $(0,1,1,0)$ (starting from the top left node, moving clockwise). 
%In 
% It corresponds to the $d = (0,1,1,0)$ (starting from the top left node, moving clockwise) representation, meaning, the $2_0$ and $2_1$ nodes have $U(1)$ gauge group each, and the other two are empty. 
That subquiver has only two fields, $w_2^{(10)}$ and $w_2^{(10)}$, subject to the F-term and D-term conditions with the choice \eqref{eq:orderingtheta}
\begin{align}
    \left(w_2^{(01)}\right)^{p}\,\left(w_2^{(10)}\right)^p&=0\\
    |w_2^{(01)}|^2-|w_2^{(10)}|^2&=\theta_{R1}-\theta_{R0}>0\,.
\end{align}
This reduces the moduli space to $\mathbb{C}[w_2^{(10)}]/\left(\left(w_2^{(10)}\right)^p\right)$ as expected.

Having established this, we easily deduce that this state is a gauge singlet by noting that the dimension vector $(0,1,1,0)$ has Dirac pairing zero w.r.t. all other possible representations, since it 
%has zero net in- and out-going arrows. In other words, 
it sits in the kernel of the antisymmetric part of the adjacency matrix \eqref{eq:admattotant}.

% In the affine orbifold coordinates $(u, v, w, z, \lambda)$ of section \ref{sec:pagodinaaffine}, this is found in the patch $w \neq 0$. By gauge-fixing that to one, we found a Pagoda given by a hypersurface
% \[
% uv = z^2-\lambda^{k-1}\,.
% \]
% Let us find the corresponding locus in quiver language. Here we want to gauge-fix $W$ in such a way that the $\mathbb{F}_2$ surface still fits inside of it:
% \begin{equation}
%      \mathbf{W} \equiv \begin{pmatrix} 0 & w^{(01)} \\ w^{(10)} & 0 \end{pmatrix}  = \begin{pmatrix}
%          0&1\\w^{(10)}&0
%      \end{pmatrix}\,.
% \end{equation}
% This Higgses the total quiver gauge group in such a way the the four nodes fuse into two nodes, resulting in the following quiver:

\section{Higgsing Pagoda-matter: Coupling frozen at \texorpdfstring{$\infty$}{infinity}}\label{sec:HiggsPagoda}
In this section, we discuss the physical significance of the class of geometries we have introduced. We claim that these give rise to 5d SCFTs that evade standard classifications, whether field-theoretic or geometric, such as those found in \cite{Intriligator:1997pq, Seiberg:1996bd, Jefferson:2018irk}, among many more works.\footnote{Note that the possibility of coupling cDV matter to higher-rank theories was already pointed out in \cite{Closset:2020scj}, Section 4.3.2, in a different example.}

The root of their peculiar physical properties lies in their inherent non-toricity. The punch line is the following: We have found geometries yielding theories that, once their matter fields acquire a VEV, \emph{do not admit any weak coupling regime}.

These geometries give rise to theories of arbitrarily high rank,\footnote{Infinite families of theories with rank $r>1$ will be introduced in Section \ref{sec:GeneralOrb}. %Later in the paper we will generalize this to higher rank.
} with an arbitrary amount of so-called \emph{Pagoda-matter} fields. We recall that these are states arising from M2-branes wrapping $\mathcal{O}(0)\oplus \mathcal{O}(-2)$-curves whose moduli are obstructed at $k+1$-th order by a superpotential. In many respect, they behave as hypermultiplets, because their moduli space is zero-dimensional \cite{Witten:1996qb}. However,
%Following Witten's argument \cite{Witten:1996qb}, these correspond to hypermultiplets because their moduli spaces are points. Moreover, 
BPS quiver analysis identifies them as indecomposable representations whose moduli spaces are fat points, i.e. $\mathbb{C}[w]/(w^k)$.
%\fdm{We are simultaneously saying two things that contradict each other: either they are hypermultiplets and their moduli space is a point, and their DT invariant is $\Omega=1$, or they are something else, their moduli space is a fat point and their DT invariant is $\Omega=k$ (our case). I would entirely delete the sentence ``following Witten's argument". Also after I would suggest ``These uncharged hypers'' --> this uncharged matter}
These uncharged hypers have a striking physical effect: their presence allows for a mechanism that freezes the gauge couplings to infinite value.

To explain this, we revisit the relation between the geometry and the effective field theory. Although we presented our geometries as $\mathbb{Z}_2$-orbifolds of the Pagoda, they can equivalently be viewed as a deformation of the \emph{local} $\mathbb{F}_2$ Calabi-Yau threefold, whereby the quivers are the same, but the superpotentials are deformations as follows:
\begin{eqnarray}
    W = W_{\text{local }\mathbb{F}_2}+W_k\,,
\end{eqnarray}
with the first term being the standard one for the toric variety

\begin{align} \label{eq:orbifolded_pagoda_sum}
W_{\text{local }\mathbb{F}_2} &= \text{Tr} \Bigl( \boldsymbol{\alpha}^{(00)} \cdot \boldsymbol{\beta}^{(01)} \, w_1^{(10)} \Bigr) 
                    + \text{Tr} \Bigl( \boldsymbol{\alpha}^{(11)} \cdot \boldsymbol{\beta}^{(10)} \, w_1^{(01)} \Bigr) \\
                   &+ \text{Tr} \Bigl( \boldsymbol{\beta}^{(10)} \cdot \boldsymbol{\alpha}^{(00)} \, w_2^{(01)} \Bigr) 
                    + \text{Tr} \Bigl( \boldsymbol{\beta}^{(01)} \cdot \boldsymbol{\alpha}^{(11)} \, w_2^{(10)} \Bigr)
\end{align}
and the second one being the order $k+1$ deformation:
\[
W_k =  \frac{2}{k+1}  \text{Tr} \Bigl( \underbrace{w_1^{(0, 1)} w_1^{(1, 0)} \dots w_1^{(1, 0)}}_{k+1 \text{ terms}} \Bigr) \\
                   - \frac{2}{k+1} \text{Tr} \Bigl( \underbrace{w_2^{(0, 1)} w_2^{(1, 0)} \dots w_2^{(1, 0)}}_{k+1 \text{ terms}} \Bigr) \:.
\]
The effective gauge coupling of a 5d theory on its Coulomb branch (parametrized by the scalar VEV $\phi$) is given by the second derivative of the prepotential $\mathcal{F}(\phi)$. For the local $\mathbb{F}_2$ theory, the one-loop exact effective coupling is \cite{Morrison:1996xf,Seiberg:1996bd}:
\begin{equation}
\label{eq:effective_coupling}
    \frac{1}{g_{eff}^2(\phi)} = \frac{\partial^2 \mathcal{F}}{\partial \phi^2} = 2m_0 + 8|\phi| \,.
\end{equation}
In a standard local $\mathbb{F}_2$ theory, $m_0$ is a tunable parameter. One can take the weak coupling limit by making the base curve very large ($m_0 \to \infty$), which decouples the instanton particles and recovers a perturbative $SU(2)$ gauge theory.
For the local $\mathbb{F}_2$ theory, we identify the lhs of \eqref{eq:effective_coupling} with twice the volume of the curve in $\mathbb{F}_2$ of self-intersection $+2$. Consequently, the geometry dictates that the effective volume of the base is shifted by the fiber volume proportional to the twist $n=2$ of the Hirzebruch surface \cite{Intriligator:1997pq}. Here, the two terms have distinct geometric interpretations:
\begin{itemize}
    \item $2\phi$ is the volume of the fiber $\mathbb{P}^1$ of the surface.
    \item $m_0$ is proportional to the inverse bare coupling squared ($m_0=\frac{1}{2g_{YM,0}^2}$). Geometrically, it is the volume of the base $(-2)$-curve of the $\mathbb{F}_2$ surface.
\end{itemize}

However, the "Pagoda matter" changes this picture drastically once we activate the further normalizable deformations corresponding to giving VEVs to these fields.
Turning on these deformations obstructs the K\"ahler modulus of the $\mathbb{F}_2$ base curve:
\begin{equation}
    m_0 \longrightarrow 0 \,.
\end{equation}
Substituting this back into Eq.~\eqref{eq:effective_coupling}, the effective coupling in this Higgsed phase becomes simply:
\begin{equation}
    \frac{1}{g_{eff}^2(\phi)} = 8|\phi| \,.
\end{equation}
This has a profound consequence. In standard theories, one can keep $\phi$ small (near the SCFT point) while making $g_{eff}$ small by dialing $m_0$ to be large. In the Higgsed Pagoda theories, this "buffer" is removed. As we approach the origin of the Coulomb branch ($\phi \to 0$), the effective coupling $g_{eff}$ diverges immediately and unavoidably.

Thus, the activation of Pagoda matter VEVs freezes the bare coupling $m_0$ to zero. This implies that in this branch of the moduli space, there is no weakly coupled regime: the theory is intrinsically strongly coupled at the origin, and cannot be continuously deformed to a perturbative non-Abelian gauge theory.

\section{Physical Interpretation: Non-constant flavor background}\label{sec:NonConstantFlavor}\label{sec:Physics}

In this section, we provide a string-theoretic interpretation of the superpotential deformation $W_k$ introduced in Eq.~\eqref{eq:W_orbifold}. This is a rather speculative section, based on upcoming work \cite{Collinucci:D2branes} by  some of the present authors, building on a progression of ideas written in \cite{Collinucci:2016hpz, Collinucci:2017bwv, Collinucci:2022rii,Moleti:2024skd}.
We propose that the Pagoda theories should be viewed as deformations of the standard toric orbifold theories, such as the conifold or $\mathbb{C}^2/\mathbb{Z}_2 \times \mathbb{C}$, or in our main case, local $\mathbb{F}_2$, where the flavor symmetry parameters are promoted to \emph{position-dependent background fields} for the $SU(2)$ global symmetry of the original 5d SCFT.

\subsection{The 5d Perspective}
The M-theory interpretation of this new phenomenon is the following:
The toric parent geometry $K_{\mathbb{F}_2}$, contains a non-compact curve of $A_1$ singularities extending along the $w$-axis. In M-theory, this singular locus supports 7d degrees of freedom (associated with vanishing 2-cycles) propagating along the entire $w$-line. This is the background $SU(2)$ vector multiplet.

The Pagoda geometry is defined by the deformation $uv - z^2 = -w^{2k}$.
Geometrically, this describes an $A_1$ fiber whose deformation parameter $\mu$ varies over the base $\mathbb{C}_w$ as $\mu(w) \sim w^{2k}$.
For any $w \neq 0$, the deformation is non-zero, the $A_1$ singularity is smoothed out, and the associated 2-cycles have finite volume. Consequently, the M2-branes wrapping these cycles become massive and non-BPS.

The physical effect of the term $w^{2k}$ is therefore to \emph{localize} the massless degrees of freedom. Unlike the toric parent, where states can propagate freely along the $w$-direction, the Pagoda geometry generates a "geometric potential" that confines the singular sector to the origin $w=0$. The \emph{Pagoda matter} consists of these localized states, trapped by the geometry itself.
The background $SU(2)$ vector multiplet contains three real adjoint scalars. We pair two into a complex field $\Phi$ and keep the third, $\varphi_3$, real. Their VEVs correspond to the geometric moduli:
\begin{align}
    \langle \Phi \rangle \quad &\longleftrightarrow \quad \text{Complex Structure (Smoothing)} \,, \\
    \langle \varphi_3 \rangle \quad &\longleftrightarrow \quad \text{K\"ahler Form (Resolution)} \,.
\end{align}
The Pagoda geometry is defined by the fixed background profile via the following spectral equation:
\begin{equation}
    uv = \det\left(\mathbb{I}_2\cdot z-\langle \Phi \rangle \right)
\end{equation}
for the following choice of background:
\begin{equation} \label{eq:backroundHiggs_pagoda}
    \Phi(w) = \begin{pmatrix} w^k & 0 \\ 0 & -w^k \end{pmatrix}.
\end{equation}
A background in $\Phi$ generates a potential for $\varphi_3$ of the form
\[V \sim |[\langle \Phi(w) \rangle, \varphi_3]|^2\]
A constant background for $\varphi_3$ corresponds to a real mass deformation of the 5d SCFT, i.e. $\langle \varphi_2 \rangle \sim m_0$. For the Pagoda background, we see that there is only one available background
\[
\langle\varphi_3 \rangle = m_0\cdot \,\begin{pmatrix}
    1&0\\0&-1
\end{pmatrix}\,.
\]
Now, following the analysis in \cite{Collinucci:2021ofd,Collinucci:2021ofd}, we know that $\Phi$ for the Pagoda has $k$ 5d dynamical fluctuations. In an appropriate gauge, they can be written as follows:
\[  \delta \Phi = \begin{pmatrix}
    0&1\\P_{k-1}(w)&0
\end{pmatrix}\,,\]
where $P_{k-1}(w)$ is an arbitrary polynomial of degree $k-1$, with $k$ coefficients. These complex structure moduli coefficients are the projection of the full Higgs branch coordinates w.r.t. their $\left(\mathbb{C}^*\right)^k$ coordinates, which are realized as Wilson lines of the 11d supergravity $C_3$-form. The consequence of giving a vev to these deformations is that our background becomes:
\begin{equation}
\langle \Phi_{\rm new}(w)\rangle = \begin{pmatrix}
    w^k&1\\P_{k-1}(w)&-w^k
\end{pmatrix}
\end{equation}
and as a result the spectral equation changes drastically:
\begin{align}
    uv &= \det\left(\mathbb{I}_2\cdot z-\langle \Phi \rangle \right)\\
        &= z^2-w^{2 k}+P_{k-1}(w)\,.
\end{align}
These are precisely the dynamical deformations according to the analysis of \cite{Shapere:1999xr}. Note now, that we can no longer find a constant $\varphi_3$ that commutes with it, i.e.
\[
[\langle \Phi_{\rm new}(w) \rangle, \varphi_3] \neq 0\,.
\]
Geometrically, this tells us that the small resolution of the CY threefold (which corresponds to blow-up the base curve of the $\mathbb{F}_2)$ is obstructed, as advertised.
For our orbifolded Pagodas, we actually apply this logic to the \emph{Pagodina}, i.e. the resulting order $\frac{k-1}{2}$ Pagoda geometry that results from orbifolding the $K_{\mathbb{F}_2}$. It is in that geometry that the K\"ahler obstruction gives rise to an important physical phenomenon.

\subsection{The view from D2-branes}

We can further clarify the origin of the superpotential deformation $W_k$ by considering the worldvolume theory of a spacetime-filling D2-brane probing the Calabi-Yau threefold. This perspective, based on work in progress \cite{Collinucci:D2branes}, utilizes 3d $\mathcal{N}=4$ mirror symmetry to map geometric deformations to field-theoretic interactions.

% OLD: In one mirror duality frame, this \Phi is a complex adjoint living on a stack...
% NEW: (Establishing a clear dictionary first)
The D2-brane probe admits two dual descriptions:
\begin{itemize}
    \item The Quiver Frame: A D2-brane probing an $A_1$ singularity ($uv = z^2$). The theory is an affine $A_1$ quiver gauge theory where the $SU(2)$ flavor symmetry is realized via monopole operators and dual photons.
    \item The SQED Frame: The mirror dual theory, which is $\mathcal{N}=4$ SQED with two hypermultiplets $(Q, \tilde{Q})$ coupled to a complex adjoint $\Phi$ of the $SU(2)$ flavor symmetry. Physically, this coressponds to a D2 probing a stack of two D6-branes in flat space. $\Phi$ is an adjoint of that stack. 
\end{itemize}

In this setup, the Pagoda geometry \eqref{kpagoda} is realized by fibering the $A_1$ singularity over a complex plane $\mathbb{C}_w$. At the level of the probe D2-brane 3d theory, this corresponds to make the flavor symmetry background $\Phi$ depend on the field $w$ controlling the position of the D2-brane on $\mathbb{C}_w$. In the SQED frame, this background enters the superpotential as a Yukawa-like interaction:
\begin{equation}\label{Eq:sup3dSQED}
    W_{\text{int}} = \text{Tr} \left( Q \cdot \langle\Phi(w)\rangle \cdot \tilde{Q} \right),
\end{equation}
where for the $k$-Pagoda case, $\Phi(w)$ takes the form \eqref{eq:backroundHiggs_pagoda}. 

In the quiver frame, the superpotential term \eqref{Eq:sup3dSQED} becomes a $w$-dependent complex FI term~\cite{Cachazo:2001gh},
\begin{equation}
     W_{\text{int}} \sim \frac{1}{k+1}\, w^{k+1}\,.
\end{equation}
This precisely matches the polynomial deformation terms $\frac{1}{k+1}\tr(w^{k+1})$ appearing in our BPS quivers.
On the other hand, by analyzing the linearized fluctuations of the background Higgs field \eqref{eq:backroundHiggs_pagoda} using the techniques of~\cite{Cecotti:2010bp}, one finds that—unlike in the undeformed $\mathbb{C}^2/\mathbb{Z}_2 \times \mathbb{C}$ background—the theory now supports modes localized at $w=0$. These localized degrees of freedom constitute what we refer to as \emph{Pagoda matter}. 
Hence these two observations are two sides of the same mechanism: the $\Phi$-generated superpotential deformation both encodes the polynomial terms in the quiver description and is responsible for the appearance of localized Pagoda matter at $w=0$.

\begin{figure}[th]
    \centering
    \begin{tikzpicture}[
        scale=1.2,
        lattice point/.style={circle, fill=gray!40, inner sep=0.8pt, outer sep=0pt},
        vertex/.style={circle, draw=blue!80!black, fill=blue!80!black, thick, minimum size=5pt, inner sep=0pt, outer sep=0pt},
        boundary node/.style={circle, draw=blue!80!black, fill=white, thick, minimum size=4pt, inner sep=0pt, outer sep=0pt},
        interior node/.style={circle, draw=blue!50!black, fill=blue!30, minimum size=4pt, inner sep=0pt},
        boundary line/.style={ultra thick, blue!80!black, line join=round, line cap=round},
        poly fill/.style={fill=blue!10, opacity=0.8},
        normal vector/.style={->, >=latex, red!80!black, thick}
    ]

    % --- 1. Define Coordinates (MOVED TO TOP) ---
    \coordinate (V1) at (0,0);
    \coordinate (V2) at (3,3);
    \coordinate (V3) at (1,4);
    \coordinate (V4) at (-2,2);

    % --- 2. Draw the Polygon Fill (FIRST, so it is in background) ---
    % Removed the scope and 'on background layer' key
    \fill[poly fill] (V1) -- (V2) -- (V3) -- (V4) -- cycle;

    % --- 3. The Background Grid (SECOND, so it sits on top of fill) ---
    \clip (-2.5, -1.5) rectangle (4.5, 5.5);
    \draw[very thin, gray!20] (-3,-2) grid (5, 6);
    \foreach \x in {-3,...,5} {
        \foreach \y in {-2,...,6} {
            \node[lattice point] at (\x,\y) {};
        }
    }

    % --- 4. Draw the Boundary Lines ---
    \draw[boundary line] (V1) -- (V2) -- (V3) -- (V4) -- cycle;

    % --- 5. Draw Outward Normals on Segment V1-V2 ---
    \foreach \i in {0, 1, 2} {
        \coordinate (mid) at ({\i+0.5}, {\i+0.5});
        \draw[normal vector] (mid) -- ++(-45:0.6);
    }

    % --- 6. Draw Special Points ---
    \node[vertex, label=below left:$v_1$] at (V1) {};
    \node[vertex, label=right:$v_2$] at (V2) {};
    \node[vertex, label=above:$v_3$] at (V3) {};
    \node[vertex, label=left:$v_4$] at (V4) {};

    \node[boundary node] at (1,1) {};
    \node[boundary node] at (2,2) {};
    \node[boundary node] at (-1,1) {}; 

    \foreach \p in {(0,1), (0,2), (-1,2), (0,3), (1,2), (1,3), (2,3)} {
         \node[interior node] at \p {};
    }

    \end{tikzpicture}
    
    \caption{Toric polytope for a threefold with a highlighted $SU(3)$ global symmetry.}
    \label{fig:toric_polygon}
\end{figure}
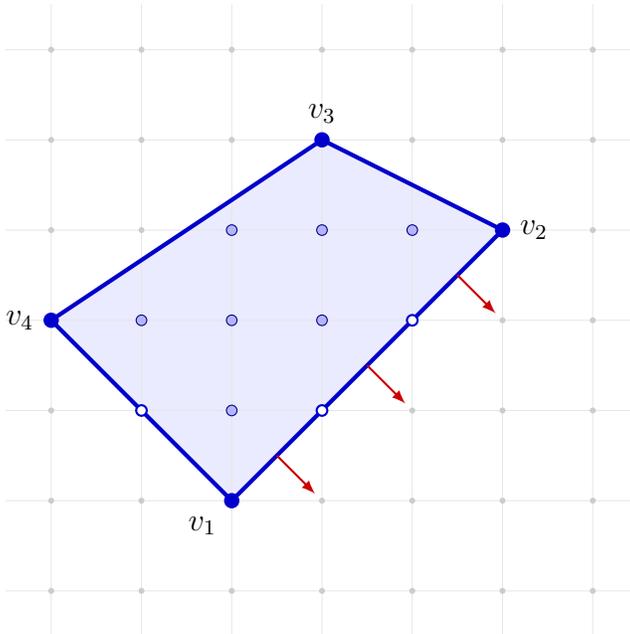

% This $\Phi$-generated term in the superpotential has two important consequences:
% \begin{enumerate}
%     \item The D2-probe sees this background as a superpotential deformation. The deformation is given by the pairing of the moment map with the primitive of the background field:
% \begin{equation}
%     \delta W \sim \int \tr(\Phi(w) \cdot \mu) \, dw \sim \int w^k \, dw \sim \frac{1}{k+1} w^{k+1} \,.
% \end{equation}
%This precisely matches the polynomial deformation terms $\frac{1}{k+1}\tr(w^{k+1})$ appearing in our BPS quivers.
%     \item By analyzing the linearized fluctuations of the background Higgs \eqref{eq:backroundHiggs_pagoda}, with the techniques of \cite{Cecotti:2010bp}, one discovers that, unlike the original $\mathbb{C}^2/\mathbb{Z}_2 \times \mathbb{C}$ background, this theory now has localized modes at $w=0$. This is \emph{Pagoda-matter}.
% \end{enumerate} 

We claim that this physical mechanism extends to all toric CY threefolds whose toric diagram has a side with points along the segment, i.e. anything that has a boundary segment as in Figure~\ref{fig:toric_polygon}. For such a theory, one can similarly introduce a position-dependent $SU(3)$ background, creating localized matter.

\begin{figure}[t]
\centering
\begin{tikzpicture}
    % Edges
    \draw (0,0) -- (1,0); \draw (0,0) -- (-1,0);
    \draw (-1,0) -- (0,-2); \draw (1,0) -- (0,-2);
    \draw (0,0) -- (0,-1); \draw (1,0) -- (0,-1); \draw (-1,0) -- (0,-1); \draw (0,-1) -- (0,-2);
    % Points
    \filldraw (-1,0) circle (2pt); \filldraw (0,0) circle (2pt);
    \filldraw (1,0) circle (2pt); \filldraw (0,-1) circle (2pt);
    \filldraw (0,-2) circle (2pt);
\end{tikzpicture}
\caption{Toric diagram local $\mathbb{F}_2$.}
\label{fig:c3mod2n}
\end{figure}
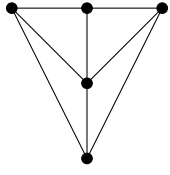

Now, we can apply this logic to local $\mathbb{F}_2$, see Figure~\ref{fig:c3mod2n}. This theory has a global $SU(2)$ symmetry. By switching on a background Higgs vev \eqref{eq:backroundHiggs_pagoda}, we claim that we end up with precisely the theory of the orbifolded Pagoda \eqref{eq:orbifolded_pagoda_sum}. The localized \emph{Pagoda-matter} multiplets are fully accounted for by the analysis of linearized fluctuations of this vev.

%%% General orbifolds %%5
% ==========================================
% SECTION 7: GENERAL ORBIFOLDS
% ==========================================
\section{More general orbifolds}\label{sec:GeneralOrb}

We can apply the same procedure to much more general orbifolds and produce a wide class of new 5d theories. The ambient space coordinates $(u,v,z,w)$ admit a larger group of symmetries that preserve the structure of the Pagoda equation \eqref{kpagoda} and its holomorphic volume form.

The matrix formalism of Section \ref{sec:McKay} is directly generalized by choosing an appropriate matrix representation of the orbifold action: fields $\phi_\pm$ of charge $\pm 1$ under an orbifold group $\mathbb{Z}_m$ are promoted to matrices of the form
{\footnotesize
\begin{equation}\label{eq:shiftm}
    \hat{\phi}_+=\left( \begin{array}{ccccc}
        0 & \phi^{(0,1)}_+ & 0 & \dots & 0 \\
        0 & 0 & \phi^{(1,2)}_+ & \ddots & 0 \\
        \vdots & \vdots & \ddots & \ddots & \vdots \\
        0 & 0 & \vdots & 0 &  \phi^{(m-1,1)}_+  \\
        \phi^{(m,0)}_+ & 0 & \cdots & 0  & 0
    \end{array} \right),\quad \hat{\phi}_-= \begin{pmatrix}
0      & 0      & \cdots & 0      & \phi^{(0,m-1)}_- \\
\phi^{(1,0)}_-& 0      & \cdots & 0      & 0 \\
0      & \phi^{(2,1)}_-& \ddots & \vdots & \vdots \\
\vdots & \ddots & \ddots & 0      & 0 \\
0      & \cdots & 0      & \phi^{(m-1,m-2)}_- & 0
\end{pmatrix},
\end{equation}
}
while neutral (charge zero) fields will be promoted to diagonal matrices. In this notation, the superpotential for all cases is given by the master trace formula derived in Section~\ref{sec:OrbifConstrPag}:
\begin{equation}
\label{eq:W_master}
W_{Orb}=\tr\left(\hat{\bs\alpha}\cdot\hat{\bs\beta} \,\hat{w}_1 \right) + \tr\left(\hat{\bs\beta}\cdot\hat{\bs\alpha}\,\hat{w}_2 \right) + \frac{1}{k+1}\tr \left(\hat{w}_1^{k+1}\right) - \frac{1}{k+1}\tr \left(\hat{w}_2^{k+1}\right).
\end{equation}
For general abelian orbifold, the D0-brane supersymmetric quantum mechanics is always described by the abelian quiver with $U(1)$ gauge group at every node. We highlight three distinct cases of interest.

% ---------------------------------------------------------
% CASE 1: Z_N
% ---------------------------------------------------------
\subsection{$\mathbb{Z}_N$ orbifolds of the Pagoda: generalization of the $\mathbb{Z}_2$ case}
We consider an orbifold group $H \cong \mathbb Z_N$, that generalizes the $\mathbb{Z}_2$ example of Section \ref{sec:pagodamodz2}. 
Its action on the affine coordinates $(u,v,z,w)$ is the following  
\begin{equation}\label{Eq:Case1Action}
\begin{array}{c|cccc}
 & u & v & z & w \\
 \hline
\mathbb{Z}_N & 1 & 1 & 1 & -1 \\
\end{array} 
\end{equation}
where we  require that $N$ divides $k+1$, i.e. $k+1 = qN$.

Like in the $\mathbb{Z}_2$ case, we can understand the geometry of the orbifolded threefold by performing a partial resolution. First, we describe the $\mathbb C^{4}/\mathbb Z_{N}$ orbifold of the ambient space via the GLSM
\begin{equation}\label{Eq:PesiC4modZN}
\begin{array}{c|ccccc}
 & u & v & z & w & \lambda \\
\hline
\mathbb{C}^* & 1 & 1 & 1 & N-1 & -N
\end{array}
\end{equation}
The geometry of the Calabi-Yau is the proper transform of \eqref{kpagoda} under this blowup: 
\begin{equation}
    \label{eq:propertrmodZn}
    u v - z^2 + w^{2k}\lambda^{2(k-q)} = 0. 
\end{equation}
Setting $\lambda = 0$, we again obtain an exceptional divisor isomorphic to a degenerate $\mathbb{F}_2$, in which the $(-2)$ curve has been shrunk. Moreover, upon restricting to the patch $w = 1$, we find a Pagodina singularity of order $k - q$.\footnote{For $N = 2$ one has $q = \tfrac{k+1}{2}$, and we recover $p = k - q = \tfrac{k-1}{2}$.}
Resolving the Pagodina curve via a small resolution, like in \eqref{Eq:PagodinaResolution}, yields an exceptional divisor isomorphic to the $\mathbb{F}_2$ surface. However, in contrast with the $\mathbb{Z}_2$ case, the resulting threefold now develops a $\mathbb{C}^2/\mathbb{Z}_{N-1}$ singularity at a point on the $\mathbb{F}_2$ fiber. This singularity is inherited from the $\mathbb{Z}_{N-1}$ orbifold singularity of the ambient space \eqref{Eq:PesiC4modZN}. Resolving it yields $N-1$ ruled surfaces, each fibered over a base isomorphic to the Pagodina curve. Altogether, the fully resolved geometry contains $r = N$ compact divisors.

\

\begin{figure}[H]
    \centering
    \def\quiverradius{2.8cm}

    \begin{subfigure}[l]{0.48\textwidth}
        \centering
        \begin{tikzpicture}[scale=0.8, >=Stealth,
            place/.style={circle,draw=black!500,fill=black!70,thick,inner sep=0pt,minimum size=6mm},
            wplace/.style={circle,draw=black!500,fill=white,thick,inner sep=0pt,minimum size=6mm}]
            \node[wplace] (N1) at (0:\quiverradius) {$1_0$};
            \node[place, text=white] (N2) at (60:\quiverradius) {$2_0$};
            \node[wplace] (N3) at (120:\quiverradius) {$1_1$};
            \node[place, text=white] (N4) at (180:\quiverradius) {$2_1$};
            \node[wplace] (N5) at (240:\quiverradius) {$1_2$};
            \node[place, text=white] (N6) at (300:\quiverradius) {$2_2$};

            \draw[<-] (N1) -- (N3);
            \draw[<-, bend left=15] (N2) to (N1); \draw[<-, bend right=15] (N2) to (N1);
            \draw[<-] (N2) -- (N4);
            \draw[<-, bend left=15] (N3) to (N2); \draw[<-, bend right=15] (N3) to (N2);
            \draw[<-] (N3) -- (N5);
            \draw[<-, bend left=15] (N4) to (N3); \draw[<-, bend right=15] (N4) to (N3);
            \draw[<-] (N4) -- (N6);
            \draw[<-] (N5) -- (N1);
            \draw[<-, bend left=15] (N5) to (N4); \draw[<-, bend right=15] (N5) to (N4);
            \draw[<-] (N6) -- (N2);
            \draw[<-, bend left=15] (N6) to (N5); \draw[<-, bend right=15] (N6) to (N5);
            \draw[<-, bend left=15] (N1) to (N6); \draw[<-, bend right=15] (N1) to (N6);
        \end{tikzpicture}
        \caption{$N=3$}
    \end{subfigure}
    \hfill
    \begin{subfigure}[r]{0.48\textwidth}
        \centering
        \begin{tikzpicture}[scale=0.8, >=Stealth,
            place/.style={circle,draw=black!500,fill=black!70,thick,inner sep=0pt,minimum size=6mm},
            wplace/.style={circle,draw=black!500,fill=white,thick,inner sep=0pt,minimum size=6mm}]
            \node[wplace] (N1) at (0:\quiverradius) {$1_0$};
            \node[place, text=white] (N2) at (45:\quiverradius) {$2_0$};
            \node[wplace] (N3) at (90:\quiverradius) {$1_1$};
            \node[place, text=white] (N4) at (135:\quiverradius) {$2_1$};
            \node[wplace] (N5) at (180:\quiverradius) {$1_2$};
            \node[place, text=white] (N6) at (225:\quiverradius) {$2_2$};
            \node[wplace] (N7) at (270:\quiverradius) {$1_3$};
            \node[place, text=white] (N8) at (315:\quiverradius) {$2_3$};

            \draw[<-] (N1) -- (N3);
            \draw[<-, bend left=15] (N1) to (N8); \draw[<-, bend right=15] (N1) to (N8);
            \draw[<-, bend left=15] (N2) to (N1); \draw[<-, bend right=15] (N2) to (N1);
            \draw[<-] (N2) -- (N4);
            \draw[<-, bend left=15] (N3) to (N2); \draw[<-, bend right=15] (N3) to (N2);
            \draw[<-] (N3) -- (N5);
            \draw[<-, bend left=15] (N4) to (N3); \draw[<-, bend right=15] (N4) to (N3);
            \draw[<-] (N4) -- (N6);
            \draw[<-, bend left=15] (N5) to (N4); \draw[<-, bend right=15] (N5) to (N4);
            \draw[<-] (N5) -- (N7);
            \draw[<-, bend left=15] (N6) to (N5); \draw[<-, bend right=15] (N6) to (N5);
            \draw[<-] (N6) -- (N8);
            \draw[<-] (N7) -- (N1);
            \draw[<-, bend left=15] (N7) to (N6); \draw[<-, bend right=15] (N7) to (N6);
            \draw[<-] (N8) -- (N2);
            \draw[<-, bend left=15] (N8) to (N7); \draw[<-, bend right=15] (N8) to (N7);
        \end{tikzpicture}
        \caption{$N=4$}
    \end{subfigure}

    \caption{
    Antisymmetric BPS quivers for $\mathbb{Z}_3$ and $\mathbb{Z}_4$ orbifolds \eqref{eq:ZNorbifPagoda}.    }
    \label{fig:case1_quivers}
\end{figure}
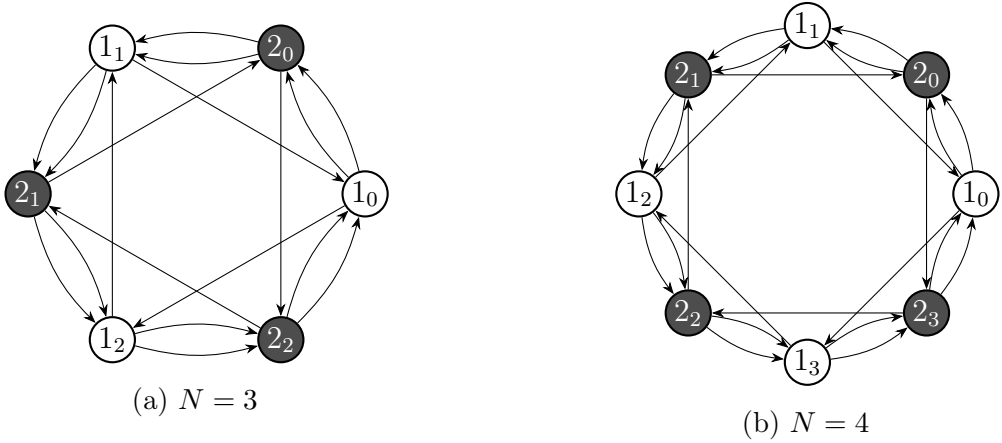

We now study the orbifold action on the quiver. The induced action on the quiver maps follows from \eqref{Eq:PesiC4modZN}:
\begin{equation}\label{eq:ZNorbifPagoda}
\begin{array}{c|cccccc}
 & \alpha_1 & \alpha_2 & \beta_1 & \beta_2 & w_1 & w_2\\
 \hline
\mathbb{Z}_N & 0 & 0 & 1 & 1 & -1 & -1 \\
\end{array}
\end{equation}
Using the matrix formalism, we can explicitly determine the adjacency matrix $A_{ab}$ of the resulting BPS quiver.\footnote{Results on the defect group of such theories will appear in \cite{Shani+}.} From it, one directly reads both the rank $r$ of the Coulomb branch and the dimension $f$ of the flavour lattice of BPS charges:
\begin{equation}\label{eq:admattotant}
A_{ab} =
\begin{cases}
2, & a=b+1 \!\!\!\!\pmod N,\\
1, & a=b-1 \!\!\!\!\pmod N,\\
0, & \text{otherwise},
\end{cases}
\qquad
r = N-1,\qquad f = 2 \, .
\end{equation}
The matrix representatives $\hat{\boldsymbol{\alpha}}$ are diagonal, while $\hat{\boldsymbol{\beta}}$ and $\hat{\boldsymbol{w}}$, that connect nodes $i\rightarrow i\pm 1$, have the form \eqref{eq:shiftm}. The corresponding quivers for $N=3,4$ are shown in Figure \ref{fig:case1_quivers}.

% ---------------------------------------------------------
% CASE 2: Z_m
% ---------------------------------------------------------
\subsection{$\mathbb{Z}_m$ orbifolds leading to Conformal Matter}
We now choose an orbifold that acts on $u$ and $v$ with opposite phases, leaving $z$ and $w$ invariant:
\begin{equation}\label{Eq:Case2Action}
\begin{array}{c|cccc}
 & u & v & z & w \\
 \hline
\mathbb{Z}_m & 1 & -1 & 0 & 0 \\
\end{array}
\end{equation}
This action is valid for any order $k$ of the Pagoda.

To understand the geometry, let us first consider the case $m=2$. The additional $\mathbb{C}^2/\mathbb{Z}_2$ singularity created by the orbifold can be resolved by passing to the hypersurface
\begin{equation}
    u\,e\,v = z^2 - w^{2k} \qquad\subset\qquad 
    \begin{array}{c|ccccc}
     & u & e & v & z & w \\
     \hline
    \mathbb{C}^\ast & 1 & -2 & 1 & 0 & 0 \\
    \end{array}
\end{equation}
There are two Pagoda singularities of order $k$, located at the two poles of the blown-up~$\mathbb{P}^1$. 
For generic $m$, the situation is analogous: resolving the ambient $\mathbb{C}^2/\mathbb{Z}_m$ singularity yields $m$ Pagoda singularities, located at the $m-2$ intersections of the exceptional $\mathbb{P}^1$’s and at one point on each external $\mathbb{P}^1$. There are no exceptional divisors. The resulting 5d theory can be regarded as a new example of 5d conformal matter \cite{DeMarco:2023irn}. 

\

\begin{figure}[h]
    \centering
    \def\quiverradius{2.8cm}

    \begin{subfigure}[l]{0.48\textwidth}
        \centering
        \begin{tikzpicture}[scale=0.8, >=Stealth,
            place/.style={circle,draw=black!500,fill=black!70,thick,inner sep=0pt,minimum size=6mm},
            wplace/.style={circle,draw=black!500,fill=white,thick,inner sep=0pt,minimum size=6mm}]
            \node[wplace] (N1) at (0:\quiverradius) {$1_0$};
            \node[place, text=white] (N2) at (60:\quiverradius) {$2_0$};
            \node[wplace] (N3) at (120:\quiverradius) {$1_1$};
            \node[place, text=white] (N4) at (180:\quiverradius) {$2_1$};
            \node[wplace] (N5) at (240:\quiverradius) {$1_2$};
            \node[place, text=white] (N6) at (300:\quiverradius) {$2_2$};

            \draw[->] (N1) -- (N2); \draw[->] (N1) -- (N4);
            \draw[->] (N2) -- (N1); \draw[->] (N2) -- (N5);
            \draw[->] (N3) -- (N4); \draw[->] (N3) -- (N6);
            \draw[->] (N4) -- (N1); \draw[->] (N4) -- (N3);
            \draw[->] (N5) -- (N2); \draw[->] (N5) -- (N6);
            \draw[->] (N6) -- (N3); \draw[->] (N6) -- (N5);

            % Self loops
            \draw[->, looseness=6] (N1) to[out=330,in=30] (N1);
            \draw[->, looseness=6] (N2) to[out=90,in=30] (N2);
            \draw[->, looseness=6] (N3) to[out=150,in=90] (N3);
            \draw[->, looseness=6] (N4) to[out=210,in=150] (N4);
            \draw[->, looseness=6] (N5) to[out=270,in=210] (N5);
            \draw[->, looseness=6] (N6) to[out=330,in=270] (N6);
        \end{tikzpicture}
        \caption{$m=3$}
    \end{subfigure}
    \hfill
    \begin{subfigure}[r]{0.50\textwidth}
        \centering
        \begin{tikzpicture}[scale=0.8, >=Stealth,
            place/.style={circle,draw=black!500,fill=black!70,thick,inner sep=0pt,minimum size=6mm},
            wplace/.style={circle,draw=black!500,fill=white,thick,inner sep=0pt,minimum size=6mm}]
            \node[wplace] (N1) at (0:\quiverradius) {$1_0$};
            \node[place, text=white] (N2) at (45:\quiverradius) {$2_0$};
            \node[wplace] (N3) at (90:\quiverradius) {$1_1$};
            \node[place, text=white] (N4) at (135:\quiverradius) {$2_1$};
            \node[wplace] (N5) at (180:\quiverradius) {$1_2$};
            \node[place, text=white] (N6) at (225:\quiverradius) {$2_2$};
            \node[wplace] (N7) at (270:\quiverradius) {$1_3$};
            \node[place, text=white] (N8) at (315:\quiverradius) {$2_3$};

            \draw[->] (N1) -- (N2); \draw[->] (N1) -- (N4);
            \draw[->] (N2) -- (N1); \draw[->] (N2) -- (N7);
            \draw[->] (N3) -- (N4); \draw[->] (N3) -- (N6);
            \draw[->] (N4) -- (N1); \draw[->] (N4) -- (N3);
            \draw[->] (N5) -- (N6); \draw[->] (N5) -- (N8);
            \draw[->] (N6) -- (N3); \draw[->] (N6) -- (N5);
            \draw[->] (N7) -- (N2); \draw[->] (N7) -- (N8);
            \draw[->] (N8) -- (N5); \draw[->] (N8) -- (N7);
        \end{tikzpicture}
        \caption{$m=4$}
    \end{subfigure}

    \caption{Symmetric BPS quivers for  $\mathbb{Z}_3$ and $\mathbb{Z}_4$ orbifolds \eqref{eq:ZmOrbifPagoda}.}
    \label{fig:case2_quivers}
\end{figure}
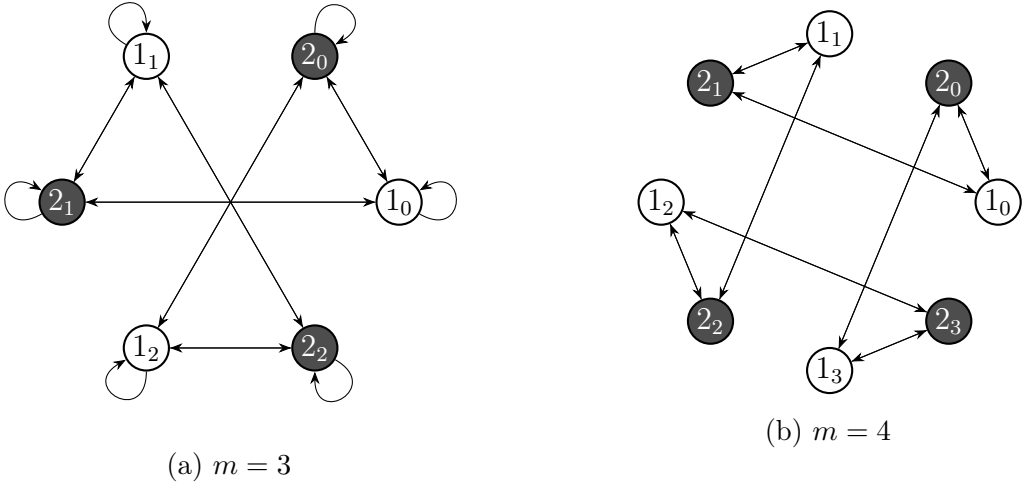

As regard the orbifold action on the Pagoda quiver, we have:
\begin{equation}\label{eq:ZmOrbifPagoda}
\begin{array}{c|cccccc}
 & \alpha_1 & \alpha_2 & \beta_1 & \beta_2 & w_1 & w_2\\
 \hline
\mathbb{Z}_m & -1 & 0 & 0 & 1 & 0 & 0 \\
\end{array}
\end{equation}
The quotient has the following adjacency matrix and values for $r$ and $f$:
\begin{equation}
A_{ab}:= \begin{cases}
        1, \quad a=b\,\text{mod} \,2m\\ 1,\quad  a=b-1\,\text{mod} \,2m\\ 1, \quad a=b+1\,\text{mod} \,2m
\\0, \quad \quad \quad\,\quad \text{otherwise}.
    \end{cases}\qquad\qquad r=0, \quad f=2m \:.
\end{equation}
The matrix fields $\hat{\alpha}_2$, $\hat{\beta}_1$, and $\hat{w}_i$ are diagonal, while $\hat{\alpha}_1$ and $\hat{\beta}_2$ have the form \eqref{eq:shiftm}.
The quivers for $m=3,4$ are in Figure~\ref{fig:case2_quivers}.

% ---------------------------------------------------------
% CASE 3: Z_m x Z_N
% ---------------------------------------------------------
\subsection{The General Abelian Orbifold $\mathbb{Z}_m \times \mathbb{Z}_N$}
Finally, we combine the actions \eqref{Eq:Case1Action} and \eqref{Eq:Case2Action}:
\begin{equation}
\begin{array}{c|cccc}
 & u & v & z & w \\
 \hline
\mathbb{Z}_m & 1 & -1 & 0 & 0 \\
\mathbb{Z}_N & 1 & 1 & 1 & -1
\end{array}
\end{equation}
The induced weights on the quiver fields are:
\begin{equation}
\begin{array}{c|cccccc}
 & \alpha_1 & \alpha_2 & \beta_1 & \beta_2 & w_1 & w_2\\
 \hline
\mathbb{Z}_m & -1 & 0 & 0 & 1 & 0 & 0 \\
\mathbb{Z}_N & 1 & 0 & 1 & 0 & -1 & -1 
\end{array}
\end{equation}
In the quotient, we have the $2mN\times 2mN$ adjacency matrix, $r$ and $f$ given by
\begin{equation}
A_{ab}:= \begin{cases}
        1, \quad a=b+2m\,\text{mod} \,2mN\\ 1,\quad  a=b+2m-1\,\text{mod} \,2mN\\ 1,\quad a=b+1\,\text{mod} \,2mN
\\0, \quad \quad \quad\,\quad \text{otherwise}
    \end{cases}\qquad 
    \begin{array}{l}
         r=mN-f/2,  \\ \\
         f=2m+ N-2 +GCD[N,2m] \:.\\ 
    \end{array}
\end{equation}
We can express the new maps as $N \times N$ block matrices, where each entry is itself an $m \times m$ matrix acting on the secondary orbifold index. 

\begin{figure}[H]
\centering

% ===== Global quiver radius =====
\def\quiverradius{3.0cm}

\begin{tikzpicture}[>=Stealth]

% ================= LEFT: 8-node quiver =================
\begin{scope}[
  place/.style={circle,draw=black,fill=black!70,thick,minimum size=6mm},
  wplace/.style={circle,draw=black,fill=white,thick,minimum size=6mm}
]
\node[wplace] (L1) at (  0:\quiverradius) {$1_0$};
\node[place,text=white] (L2) at ( 45:\quiverradius) {$2_0$};
\node[wplace] (L3) at ( 90:\quiverradius) {$1_1$};
\node[place,text=white] (L4) at (135:\quiverradius) {$2_1$};
\node[wplace] (L5) at (180:\quiverradius) {$1_2$};
\node[place,text=white] (L6) at (225:\quiverradius) {$2_2$};
\node[wplace] (L7) at (270:\quiverradius) {$1_3$};
\node[place,text=white] (L8) at (315:\quiverradius) {$2_3$};

\draw[->] (L1)--(L2);\draw[->] (L1)--(L4);\draw[->] (L1)--(L5);
\draw[->] (L2)--(L3);\draw[->] (L2)--(L5);\draw[->] (L2)--(L6);
\draw[->] (L3)--(L6);\draw[->] (L3)--(L7);\draw[->] (L3)--(L8);
\draw[->] (L4)--(L3);\draw[->] (L4)--(L5);\draw[->] (L4)--(L8);
\draw[->] (L5)--(L1);\draw[->] (L5)--(L6);\draw[->] (L5)--(L8);
\draw[->] (L6)--(L1);\draw[->] (L6)--(L2);\draw[->](L6)--(L7);
\draw[->] (L7)--(L2);\draw[->] (L7)--(L3);\draw[->] (L7)--(L4);
\draw[->] (L8)--(L1);\draw[->] (L8)--(L4);\draw[->] (L8)--(L7);

\node at (0,-4.6) {(a) $m=2,\, N=2$};
\end{scope}

% ================= RIGHT: 12-node quiver =================
\begin{scope}[xshift=8.2cm,
  node/.style={circle, draw, minimum size=6mm},
  odd/.style={fill=white},
  even/.style={fill=gray!70!black, text=white}
]

\node[node,odd]  (R1)  at (  0:\quiverradius) {$1_0$};
\node[node,even] (R2)  at ( 30:\quiverradius) {$2_0$};
\node[node,odd]  (R3)  at ( 60:\quiverradius) {$1_1$};
\node[node,even] (R4)  at ( 90:\quiverradius) {$2_1$};
\node[node,odd]  (R5)  at (120:\quiverradius) {$1_2$};
\node[node,even] (R6)  at (150:\quiverradius) {$2_2$};
\node[node,odd]  (R7)  at (180:\quiverradius) {$1_3$};
\node[node,even] (R8)  at (210:\quiverradius) {$2_3$};
\node[node,odd]  (R9)  at (240:\quiverradius) {$1_4$};
\node[node,even] (R10) at (270:\quiverradius) {$2_4$};
\node[node,odd]  (R11) at (300:\quiverradius) {$1_5$};
\node[node,even] (R12) at (330:\quiverradius) {$2_5$};

\draw[->] (R1)--(R2);\draw[->] (R1)--(R7);\draw[->](R1)--(R10);
\draw[->] (R2)--(R5);\draw[->] (R2)--(R7);\draw[->](R2)--(R8);
\draw[->] (R3)--(R4);\draw[->] (R3)--(R6);\draw[->](R3)--(R9);
\draw[->] (R4)--(R1);\draw[->] (R4)--(R9);\draw[->](R4)--(R10);
\draw[->] (R5)--(R8);\draw[->] (R5)--(R11);\draw[->](R5)--(R12);
\draw[->] (R6)--(R5);\draw[->] (R6)--(R9);\draw[->] (R6)--(R12);
\draw[->] (R7)--(R1);\draw[->] (R7)--(R4);\draw[->] (R7)--(R8);
\draw[->] (R8)--(R1);\draw[->] (R8)--(R2);\draw[->] (R8)--(R11);
\draw[->] (R9)--(R3);\draw[->](R9)--(R10);\draw[->] (R9)--(R12);
\draw[->] (R10)--(R3);\draw[->](R10)--(R4);\draw[->] (R10)--(R7);
\draw[->] (R11)--(R2);\draw[->] (R11)--(R5);\draw[->] (R11)--(R6);
\draw[->] (R12)--(R3);\draw[->] (R12)--(R6);\draw[->](R12)--(R11);

\node at (0,-4.6) {(b) $m=3,\, N=2$};
\end{scope}

\end{tikzpicture}
\caption{BPS quivers for the combined orbifold cases.}
\label{fig:case3_quivers}
\end{figure}
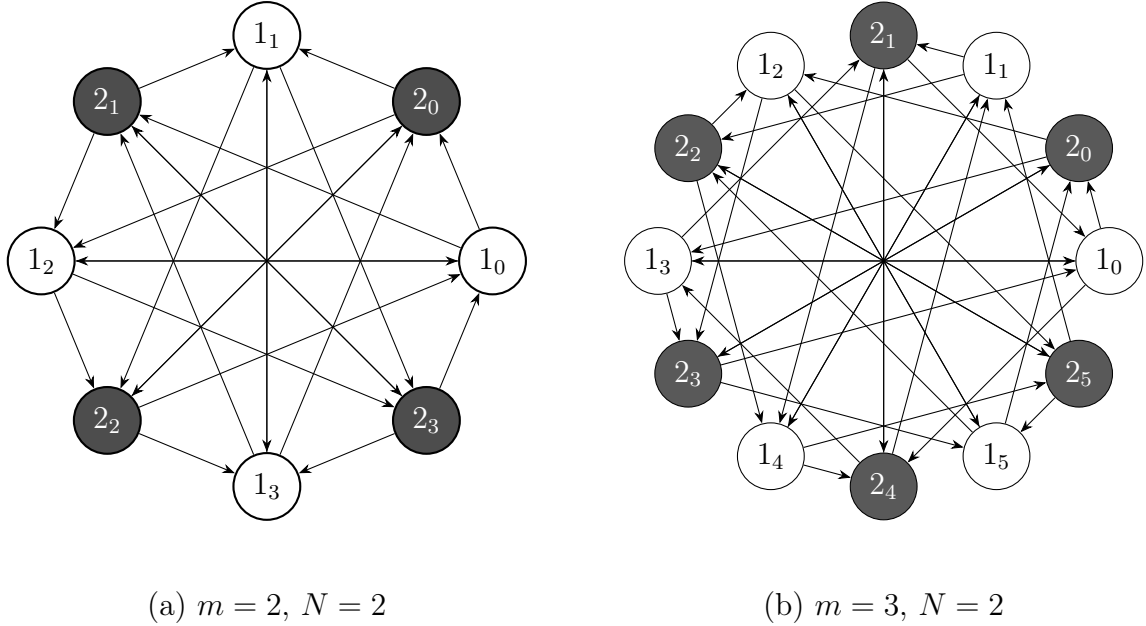

The resulting quivers for $N=m=2$ and $N=m=3$ are shown in Figure~\ref{fig:case3_quivers}.

\subsection{Orbifold of Reid Pagodas as deformations of toric orbifolds}

We conclude this section with a unifying remark. As we have said above, we can view the Reid Pagoda \eqref{kpagoda} as a deformation of the toric orbifold $Y_{\text{toric}} \equiv \mathbb C \times \mathbb C^2/\mathbb Z_2$ by the term $w^{2k}$.\footnote{For a discussion of abelian and non-abelian orbifolds of $\mathbb C^3$ and their BPS quiver we refer to \cite{DelZotto:2022fnw}.}
Since all the orbifold actions discussed above preserve $Y_{\text{toric}}$, we can commute the operations:
\begin{equation}
    X = \frac{\text{Deform}\left(Y_{\text{toric}}\right)}{H}=\text{Deform}\left( \frac{Y_{\text{toric}}}{H} \right).
\end{equation}
The geometry $\frac{Y_{\text{toric}}}{H}$ is a toric Calabi–Yau orbifold. Our final geometries $X$ are non-toric deformations of these spaces. The deformation terms do not modify the compact divisors; however, they crucially lift some of the non-compact lines of singularities present in the toric model, leaving behind discrete “Pagodina” curves that support the exotic matter.

% This relationship is summarized in the commutative diagram of Figure \ref{fig:toric-diagram}.
% \begin{figure}[H]
% \centering
% \begin{tikzcd}[row sep=3.4em, column sep=5em]
% \mathbb{C}^3 \arrow[d, "/\mathbb{Z}_2"'] \\
% \displaystyle \frac{\mathbb{C}^2}{\mathbb{Z}_2}\times \mathbb{C}
%   \arrow[r, "\mathrm{defs}"] \arrow[d, "/H"'] & Y \arrow[d, "/H"] \\
% X_{\mathrm{toric}} \arrow[r, "\mathrm{defs}"] & X
% \end{tikzcd}
% \caption{Commutative diagram relating the toric quotient and the non-toric deformation.}
% \label{fig:toric-diagram}
% \end{figure}

\section*{Acknowledgments}
The authors thank  Andrea Sangiovanni for helpful discussions and relevant comments.  
The research of A.C. and M.D.M. is funded through an ARC advanced project, and further supported by IISN-Belgium (convention 4.4503.15).
A.C. is a Senior Research Associate of the F.R.S.-FNRS (Belgium). 
M.M. and R.V. acknowledge support by INFN Iniziativa Specifica ST\&FI.
The work of F.D.M. was partly supported by the RPG-2021-047 Leverhulme grant ``Extended Riemann-Hilbert Correspondence, Quantum Curves, and Mirror Symmetry ''.

\begin{appendix}
    \section{BPS quivers of 5d SCFTs}\label{app:BPSQuivers}

The abelian quivers described in Section \ref{sec:PagodaCon} are a special instance of a more general object called \emph{BPS quiver}, which describe BPS states of supersymmetric theories with eight supercharges, which for us will be 5d theories obtained by M-theory reduction on a local CY3 $X$. In this context, BPS states arise from branes wrapping holomorphic cycles in a resolution $\hat{X}$. Using type IIA language, these can be described as bound states of D0-D2-D4 branes (see \cite{Closset2022} for more details), and the conserved charges of the 5d theory can be identified with Chern classes in the even cohomology with compact support. One has
\begin{equation}
    \Gamma:=H^0_{cpt}(X,\mathbb{Z})\oplus H^2_{cpt}(X,\mathbb{Z})\oplus H^4_{cpt}(X,\mathbb{Z})\cong \mathbb{Z}^{2r+f}.
\end{equation}
The integers $r$ and $f$ are respectively the rank of the Coulomb branch and the number of conserved flavour charges including the KK momentum, identified with the D0-brane charge. In this context, nodes of the quiver are associated to so-called \emph{fractional branes} \cite{Aspinwall2006,beaujard2021vafa,Closset2022}, and any D0-D2-D4 bound state can be equivalently seen as a bound state of such objects. The BPS quiver with superpotential encodes the supersymmetric quantum mechanics describing the low-energy physics of these BPS states: if the BPS state is a bound state of $N_i$ fractional branes associated to the $i$-th node of the quiver, then the gauge group at that node will be $U(N_i)$\footnote{Mathematically, one would consider a quiver without any gauge groups as being associated to a phase of the 5d theory, and a choice of gauge group for every node would describe a \emph{quiver representation}, see for example the physically-oriented review in the book \cite{Aspinwall2009}. We will mostly refrain from using the language of quiver representations in this work.}. A central object is the antisymmetric adjacency matrix $B$:
\begin{eqnarray}
    B_{ij}=(\text{\# arrows } i\to j)-(\text{\# arrows } j\to i),
\end{eqnarray}
which represents the Dirac pairing among BPS charges. Flavour charges have by definition zero Dirac pairing with everything, so that having the matrix $B$ we can immediately identify the flavour and Coulomb charges respectively as
\begin{equation}
    \Gamma_f=\ker B,  \qquad\Gamma_c=\Gamma_c=\mathrm{coker}B,
\end{equation}
i.e.
\begin{equation}
    f=\dim\ker B,\qquad 2r=\mathrm{rk}\,B.
\end{equation}
% In particular, the $\mathrm{rk} B$ tells us how ``dynamically nontrivial'' is the physics on the Coulomb branch, and translates to a much richer BPS physics. 
Finally, as explained in \cite{DelZotto:2022fnw}, the defect group of the geometrically engineered QFT can be computed as
\begin{equation}
\text{Tor}\left(\text{coker}(B)\right) \cong \mathbb G_{e}^{(1)} \oplus \mathbb G_{m}^{(1)},
\end{equation}
with $\mathbb G_{e}^{(1)}$ the electric one-form symmetry, and $\mathbb G_{m}^{(1)}$ the 4d KK descendant of 5d magnetic two-form symmetry. 
For more details on the subject of defect groups for orbifolds, see the upcoming work \cite{Shani+}.
\end{appendix}

% --- References ---
\clearpage
\phantomsection
\addcontentsline{toc}{section}{References}
\bibliographystyle{alph}
\bibliography{cc_bib}

\end{document}